\documentclass[aps,prl,superscriptaddress,citeautoscript,twocolumn,reprint,longbibliography,floatfix]{revtex4-2}

\usepackage[T1]{fontenc}
\usepackage{lmodern}
\usepackage{graphicx}
\usepackage{amssymb}
\usepackage{amsmath}
\usepackage{mathtools}
\usepackage{cprotect}
\usepackage[italicdiff]{physics}
\usepackage{hyperref}

\usepackage{mycommands}

\usepackage{xrhelper}

\newcommand*{\myexternaldocument}[1]{
    \externaldocument[S-]{build/#1} 
    \addFileDependency{#1.tex}
    \addFileDependency{build/#1.aux} 
}

\myexternaldocument{suppl-arxiv} 

\begin{document}

\title{Vortex reversal is a precursor of confined bacterial turbulence}

\author{Daiki Nishiguchi}
\email{nishiguchi@noneq.phys.s.u-tokyo.ac.jp}
\affiliation{Department of Physics,\! The University of Tokyo,\! 7--3--1 Hongo,\! Bunkyo-ku,\! Tokyo 113--0033,\! Japan}%

\author{Sora Shiratani}
\affiliation{Department of Physics,\! The University of Tokyo,\! 7--3--1 Hongo,\! Bunkyo-ku,\! Tokyo 113--0033,\! Japan}%

\author{Kazumasa A. Takeuchi}
\affiliation{Department of Physics,\! The University of Tokyo,\! 7--3--1 Hongo,\! Bunkyo-ku,\! Tokyo 113--0033,\! Japan}%
\affiliation{Institute for Physics of Intelligence, The University of Tokyo,\! 7--3--1 Hongo,\! Bunkyo-ku,\! Tokyo 113--0033,\! Japan}%

\author{Igor S. Aranson}
\affiliation{Departments of Biomedical Engineering, Chemistry, and Mathematics, The Pennsylvania State University, University Park, Pennsylvania 16802, USA}%
\affiliation{Department of Physics,\! The University of Tokyo,\! 7--3--1 Hongo,\! Bunkyo-ku,\! Tokyo 113--0033,\! Japan}%

\date{\today}

\begin{abstract}
    Active turbulence, or chaotic self-organized collective motion, is often observed in concentrated suspensions of motile bacteria and other systems of self-propelled interacting agents.
    To date, there is no fundamental understanding of how geometrical confinement orchestrates active turbulence and alters its physical properties.
    Here, by combining large-scale experiments, computer modeling, and analytical theory, we have discovered a generic sequence of transitions occurring in bacterial suspensions confined in cylindrical wells of varying radii. With increasing the well's radius, we observed that persistent vortex motion gives way to periodic vortex reversals, four-vortex pulsations, and then well-developed active turbulence. Using computational modeling and analytical theory, we have shown that vortex reversal results from the nonlinear interaction of the first three azimuthal modes that become unstable with the radius increase. The analytical results account for our key experimental findings. To further validate our approach, we reconstructed equations of motion from experimental data. Our findings shed light on the universal properties of confined bacterial active matter and can be applied to various biological and synthetic active systems.
\end{abstract}

\maketitle

Interacting self-propelled particles, often termed active matter,
exhibit a remarkable tendency to self-organization and the onset of collective behavior.
Being intrinsically out-of-equilibrium, active matter systems exhibit a slew of collective phenomena such as the spontaneous onset of long-range order~\cite{nishiguchi2017longrange,chate2020dry,iwasawa2021algebraic,nishiguchi2023deciphering,aranson2022bacterial}, odd viscoelasticity~\cite{fruchart2023odd}, rectifications of chaotic flows~\cite{sokolov2010swimming,reichhardt2017ratchet,nishiguchi2018engineering,reinken2020organizing}, and reduction of the effective viscosity~\cite{sokolov2009reduction,martinez2020combined}.
One of the most visible manifestations of collective dynamics in active matter systems is the emergence of self-sustained
spatiotemporal chaotic flows termed active turbulence~\cite{sokolov2007concentration,sokolov2012physical,wensink2012mesoscale, dunkel2013fluid, dunkel2013minimal, qi2022emergence}.
In stark contrast to conventional Navier-Stokes turbulence, active turbulence, occurring for essentially zero Reynolds numbers, is characterized by the well-defined characteristic length scale.
In the case of bacterial turbulence, this scale corresponds to typical vortex size, which is about 40--50 $\mathrm{\mu}$m~\cite{sokolov2012physical,wensink2012mesoscale}.
The existence of the typical vortex size allows transforming bacterial motion into stable vortex arrays under geometrical confinements~\cite{wioland2013confinement,lushi2014fluid,wioland2016directed,wioland2016ferromagnetic,beppu2017geometrydriven,beppu2021edge} or in the presence of periodic obstacles~\cite{nishiguchi2018engineering,reinken2020organizing}. 

Experimental and computational studies of self-organization of bacterial and related active systems have shown that strong confinement, e.g., a cylindrical well, may suppress active turbulence and generate persistent vortex motion~\cite{wioland2013confinement,lushi2014fluid,wioland2016ferromagnetic,beppu2017geometrydriven,beppu2021edge}.
However, a fundamental question on the nature of the transition from ordered states under strong confinement to chaotic motion in unconstrained systems remains open.
Answering this question will shed light on intricate fundamental mechanisms of self-organization in a broad class of active systems under confinement.

In the context of active nematics exemplified by microtubules-motors assays, multiple experimental and numerical studies interrogated a transition from ordered quasi-stationary states to chaotic motion occurs under the confinement in channels, rings, and wells~\cite{wu2017transition, opathalage2019self,hardouin2019reconfigurable,hardouin2022active,schimming2023friction,ravnik2013confined}. The primary observation is that the instability of static nematic configuration occurs via unbinding and subsequent chaotic motion of half-integer topological defects.
In polar active systems such as bacterial suspensions, experimental investigations have been hindered by the difficulty in resolving the detailed dynamics very close to the transition point and the necessity of long-time measurements for evaluating the vortex stability.

Here, we examine the route to active turbulence by combining large-scale experiments, high-resolution numerical modeling, and analytical theory.
We focused on a well-characterized active system: suspensions of swimming bacteria~\cite{aranson2022bacterial}.
We confined the suspensions into an array of isolated cylindrical wells comparable to the size of individual vortices. We systematically varied the wells' radii to characterize the transition from stabilized vortices to bacterial turbulence.
Increasing the well radius, we have detected reversals of vortex rotation as the first instability from a stable vortex. The reversals were also captured as periodic oscillations in our numerical simulations and analytical theory, unraveling a robust fundamental mechanism for the onset of polar active turbulence. It differs from the reversals caused by viscoelasticity of the suspending fluid~\cite{liu2021viscoelastic}.
Our analysis revealed that the reversal originated from the nonlinear interaction of the three lowest azimuthal modes near the threshold of linear instability. To further validate our theoretical arguments, we reconstructed equations of motion from experiential data. Our studies indicate that the vortex reversal is a generic precursor of turbulence-like behavior in bacterial and related active systems. Our findings provide insights into how geometrical confinement orchestrates spatiotemporal organization in a broad class of active systems.

\begin{figure*}[tb]
    \centering
    \includegraphics[width=\hsize]{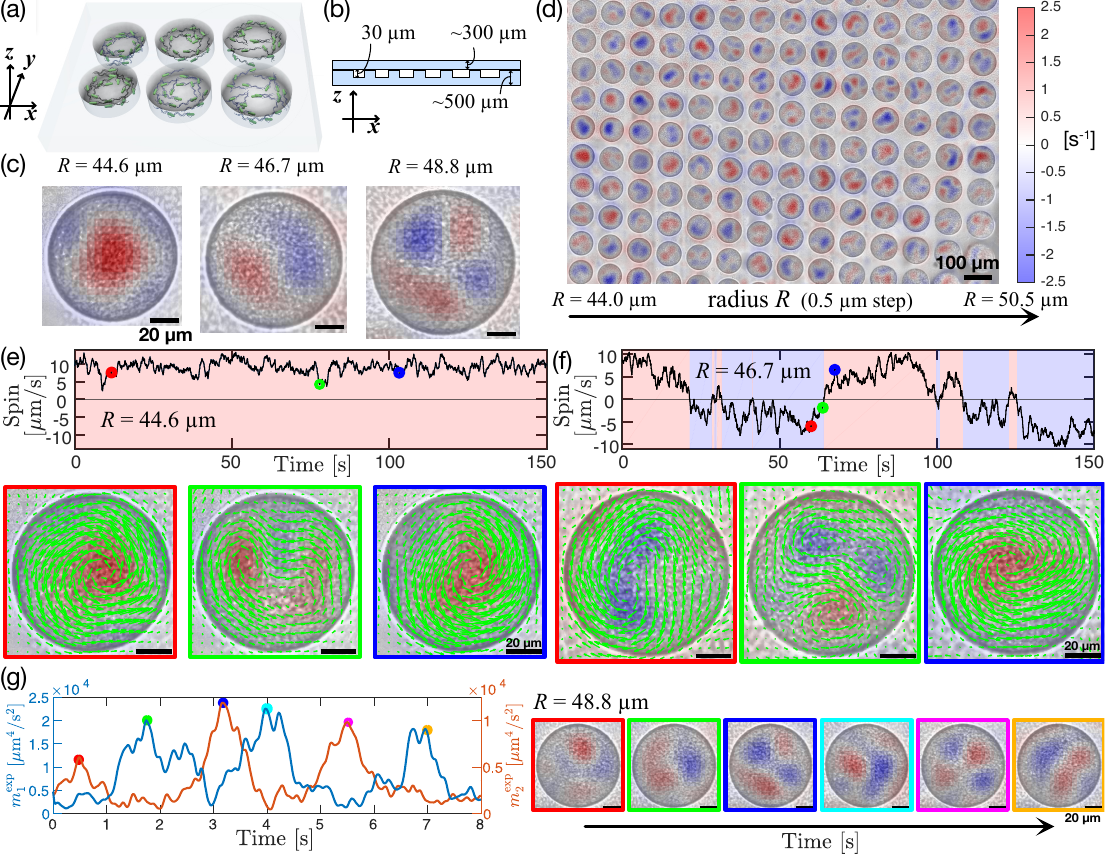}
    \caption{%
        \textbf{Transitions from a stabilized vortex to reversing vortices and a four-vortex state}.
        3D schematics (a) and side view (b) of the experimental setup.
        (c) Typical vorticity profiles of a single stabilized vortex ($R=44.6$~$\mathrm{\mu m}$), reversing vortices ($R=46.7$~$\mathrm{\mu m}$), and a four-vortex state ($R=48.8$~$\mathrm{\mu m}$). Vorticity field $\omega$ is overlaid on the experimental snapshots. The color scales of the vorticity fields in all panels are identical and are indicated by the color bar in (d). (Supplementary Movies 2--4).
        (d) Experimental snapshot overlaid with the instantaneous vorticity field. Wells with the same radius are arranged vertically, with the radius increasing from left to right. All the 119 wells within this image out of $\sim400$ wells within the whole field of view (\figref{S-figS1_WholeFiledOfView}, Supplementary Movie 1) were used for analysis, see Supplementary Note 1 and \figref{S-figS3_CWbias} for the selection criteria.
        (e,f) Time series of spins for the wells with the radii of 44.6 $\mathrm{\mu m}$ (e) and 46.7 $\mathrm{\mu m}$ (f), respectively. The instantaneous velocity and vorticity fields are shown below the time series, with the colors of the rectangle corresponding to the time points highlighted by colored circles in the time series (Supplementary Movie 2, 3, 5, 6).
        (g) Anti-phase relation of mode amplitudes $m_1^\mathrm{exp}$ and $m_2^\mathrm{exp}$ of the four-vortex state at $R=48.8$~$\mathrm{\mu m}$. The instantaneous vorticity fields are shown on the right of the time series, with the colors of the rectangle corresponding to the time points highlighted by colored circles in the time series, see Supplementary Note 4 and Supplementary Movies 3, 7.
    }\label{fig1_ExpSetup}
\end{figure*}
\subsection{Results}
\begin{figure}[tb]
    \centering
    \includegraphics[width=\hsize]{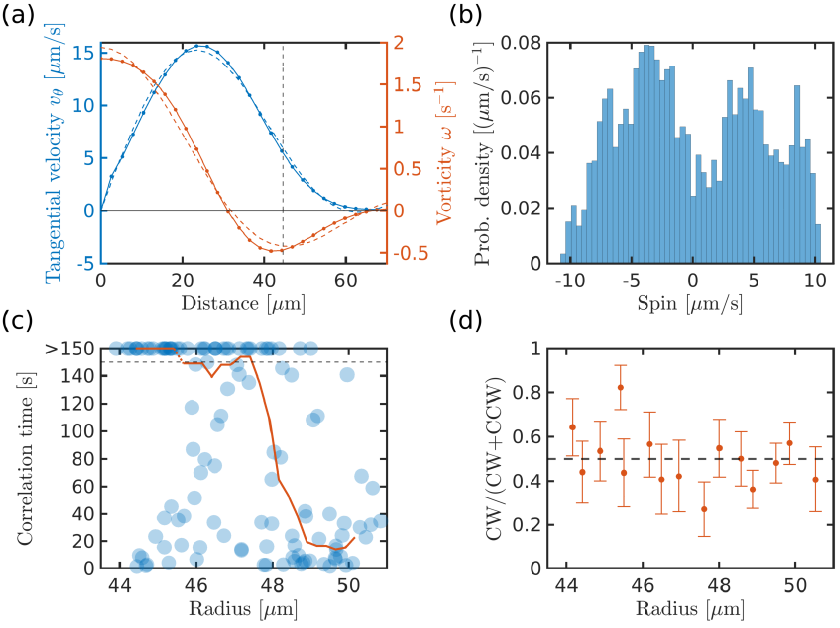}
    \caption{\textbf{Characterization of vortex states}.
        (a) Time-averaged velocity $\vb{v}$ (blue) and vorticity $\omega$ (red) profiles of a stabilized vortex shown in \figref{fig1_ExpSetup}(d). The vertical dashed line corresponds to the radius detected from the image analysis, see Supplementary Note 1. The dashed lines are fits to the analytical solutions, Eq.~(\ref{sol1}), for tangential velocity $v_\theta$ (blue) and vorticity $\omega$ (red) fields.
        (b) Spin probability density function for the reversing vortex state shown in \figref{fig1_ExpSetup}(f).
        (c) A scatter plot of the spin correlation time and the well radii detected from the image analysis. The red line represents the moving median of the scatter plot. The horizontal dashed line corresponds to the experimental duration, 150 seconds, see Supplementary Note 1 for details.
        (d) Fraction of CW rotations as a function of the well radius. Error bars are the standard errors.
    }\label{fig2_ExpData}
\end{figure}

\paragraph{Experiment.}
We conducted experiments with suspensions of swimming bacteria confined in cylindrical wells, \figref{fig1_ExpSetup}(a,b).
The height of the wells was set to 30~$\mathrm{\mu m}$, which is smaller than the typical length scale of collective motion, ensuring effectively two-dimensional dynamics within each well.
Experiments were conducted simultaneously in an array of isolated wells of different radii ($\approx400$ wells in total), see \figref{fig1_ExpSetup}(d), \figref{S-figS1_WholeFiledOfView}, and Supplementary Movies 1--7.
We observed stabilized vortices with steady rotational directions within the wells with small radii. For larger radii, the vortices exhibited a transition to unsteady configurations with reversing rotation directions. This observation is exemplified by the instantaneous vorticity field $\omega(\vb{r}, t) = \vu{z} \cdot[\nabla \times \vb{v}(\vb{r},t)]$ shown in \figref{fig1_ExpSetup}(c), where $\vu{z}$ is the unit vector in the $z$-direction. As one sees from \figref{fig1_ExpSetup}(c), the smaller wells hosted a single stabilized vortex with persistent rotation, with the velocity and vorticity profiles shown in \figref{fig1_ExpSetup}(e) and \figref{fig2_ExpData}(a).
In contrast to previous studies on bacterial suspensions confined in water-in-oil droplets~\cite{wioland2013confinement,lushi2014fluid}, we did not observe any counter-rotating edge flows, suggesting different boundary conditions for collective motion.

To quantify the vortex rotation direction, we defined a spin variable for each well as,
\begin{align}
    S_i(t) := \frac{\vu{z} \cdot
        {\displaystyle \sum_{\vb{r} \in i\text{-th well}}} (\vb{r}-\vb{r}_i) \times \vb{v}(\vb{r},t)}{\displaystyle \sum_{\vb{r} \in i\text{-th well}} |\vb{r}-\vb{r}_i|},
    \label{eq:spin}
\end{align}
where $\vb{r}_i$ is the center of the $i$-th well, and the summations run over the area of the $i$-th well.
As shown in \figref{fig1_ExpSetup}(e,f), the spins for the small wells stayed almost constant and rarely flipped their signs over time, while the spins for the larger wells persistently alternated their signs, reflecting the reversals of vortices. The spin probability distribution for such a well with reversals exhibits a bimodal distribution, indicating the presence of two states with clockwise (CW, $S_i<0$) and counterclockwise (CCW, $S_i>0$) rotations [\figref{fig2_ExpData}(b)].
Contrary to Ref.~\cite{beppu2021edge},
our setup uses two symmetric surfaces for the top and bottom to compensate for systematic bias in the rotation direction. Thus, the fraction of CW rotations as a function of the well's radius was always $\approx 0.5$, \figref{fig2_ExpData}(d). The absence of bias is crucial for characterizing the vortex reversals.
The transition from a single stabilized vortex to reversing vortices was inspected through the spin correlation time, defined as the time at which the autocorrelation function of the spins decayed to $1/e$.
The correlation time has successfully captured the transition at the radii of approximately 46 to 48 $\mathrm{\mu m}$ [\figref{fig2_ExpData}(c)]. For large wells' radii, four-vortex pulsating states were observed as well, Fig.~\ref{fig1_ExpSetup}(c).
The pulsation was characterized by the kinetic energy of azimuthal modes corresponding to $2n$ vortices within a well (see Supplementary Note 4),
\begin{gather}
    m_{n}^{\mathrm{exp}} = \int_0^R \dd r r \left| \frac{1}{2\pi}\int \dd \theta e^{-in\theta} \vb{v}(r,\theta) \right|^2.
\end{gather}
By this mode analysis, we probed anti-phase oscillation of the modes $n=1$ and $n=2$, see \figref{fig1_ExpSetup}(g).

\begin{figure*}[tb]
    \centering
    \includegraphics[width=\hsize]{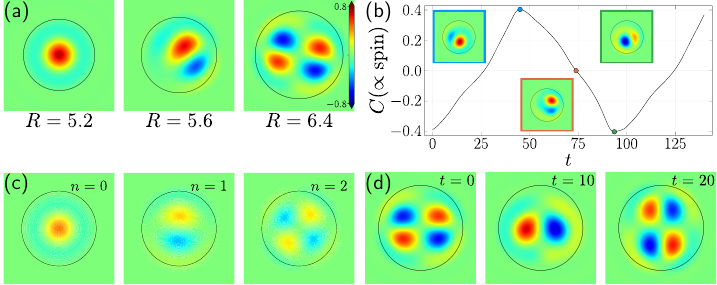}
    \caption{\textbf{Computational modeling using the TTSHE.}
        (a) Vorticity profiles obtained in the numerical simulations. Typical snapshots of a single stabilized vortex ($R=5.2$, Supplementary Movie 8), periodically reversing two-vortex state ($R=5.6$, Supplementary Movie 9), and a pulsating four-vortex state ($R=6.4$, Supplementary Movie 10) are shown. The color bar in all panels is the same.
        (b) Time series of the spin for the reversing two-vortex state ($R=5.35$, Supplementary Movie 11). The instantaneous vorticity fields are shown as insets, with the colors of the rectangle corresponding to the time points highlighted by colored circles in the time series.
        For computational convenience, instead of using the spin defined in \eqref{eq:spin}, we plot the amplitude $C$ of the zeroth azimuthal mode defined in~\eqref{omega}, because $C$ is proportional to the spin (see Supplementary Note 4).
        (c) Azimuthal mode decomposition of the instantaneous vorticity field for the reversing two-vortex state shown in the middle panel of (a) ($R=5.6$, Supplementary Movie 12), which is defined as $\int \omega\qty(r,\theta) \dd{\theta} / 2\pi$ for $n = 0$ and $\int e^{-i n \theta} \omega\qty(r,\theta) \dd{\theta} / 2\pi + \mathrm{c.c.}$ otherwise.
        (d) Snapshots of the pulsating four-vortex state ($R=6.2$, Supplementary Movie 13).
    }\label{fig3_Num}
\end{figure*}

\begin{figure*}[tb]
    \centering
    \includegraphics[width=\hsize]{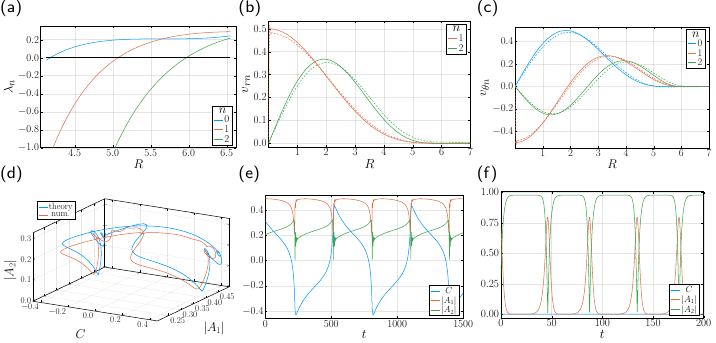}
    \caption{\textbf{Analytical results}.
        (a) The growth rates $\lambda_n$ vs radius $R$ for $n=0,1,2$.
        (b,c) Comparison of velocity profiles for the azimuthal modes with $n=0,1,2$ obtained from the linear theory (solid), Eq.~(\ref{sol1}), and the TTSHE simulations (dashed).
        (d) Comparison of trajectories in 3D phase space obtained by the solution of Eqs.~(\ref{C})-(\ref{A2}) and the TTSHE\@.
        (e,f) Amplitudes $C, A_1, A_2$ vs time obtained from
        Eqs.~(\ref{C})-(\ref{A2}) for $R=5.9$ (e) and $R=7.0$ (f), see Supplementary Movies 14, 15.
    }\label{fig4_Thm}
\end{figure*}

\paragraph{Computational modeling.}
We performed numerical simulations using a phenomenological active fluid model, the Toner-Tu-Swift-Hohenberg equation (TTSHE)~\cite{wensink2012mesoscale, dunkel2013fluid, dunkel2013minimal,reinken2020organizing, aranson2022bacterial}. The TTSHE qualitatively captures the bulk properties of polar active turbulence. It can describe the transformation of bacterial turbulence into stable vortex arrays in the presence of periodic obstacles~\cite{nishiguchi2018engineering,reinken2020organizing}.
In the vorticity representation, the dimensionless TTSHE is of the form\cite{reinken2020organizing}:
\begin{equation}
    \begin{multlined}
        \pdv{\omega}{t} + \lambda \vb{v} \vdot \grad{\omega} = a \omega - b \curl{\qty[|\vb{v}|^2 \vb{v}]} \\
        - \qty(1 + \laplacian)^2 \omega - \gamma_{\vb{v}} \curl{\qty[K\qty(\vb{r})\vb{v}]} - \gamma_{\omega} K \qty(\vb{r}) \omega,
    \end{multlined}
    \label{eq:orig2}
\end{equation}
where $\lambda$, $a$, and $b$ are constants, $K\qty(\vb{r}) \geq 0$ is a scalar field that
dampens $\vb{v}$ and $\omega$ outside the well ($K \simeq 1$) without affecting the inside ($K \simeq 0$), and $\gamma_{\vb{v},\omega} > 0$ are damping coefficients.
In this dimensionless form, the vortex characteristic size is $2\pi$.
Following Ref.~\cite{reinken2020organizing}, we adopt the parameter values
$\qty(\lambda,a,b,\gamma_{\vb{v}},\gamma_{\omega})=\qty(9,0.5,1.6,40,4)$ and impose three boundary conditions on well's wall,
\begin{equation}
    \vb{v} = \vb{0},\;\omega =0 \;\; \mbox{at} \;\; r=R.
    \label{bc}
\end{equation}
Compared with the Navier-Stokes equation, the extra boundary condition $\omega=0$ is imposed due to the higher-order differential operator ($\nabla^4$) in Eq.~(\ref{eq:orig2}).
We solved Eq.~(\ref{eq:orig2}) with the above boundary conditions in two dimensions by the pseudospectral method, see Methods.

Our simulations successfully reproduced the entire sequence of transitions observed in experiments, \figref{fig3_Num}(a) and Supplementary Movies 8--13.
We have found a single stable vortex for small radii. As the radius increases, the vortex becomes destabilized and yields a periodically reversing two-vortex state, see \figref{fig3_Num}(b,c).
It was demonstrated by the time series of the spin variable, see \figref{fig3_Num}(b). Increasing the radius, the reversing two-vortex state transforms into a pulsating four-vortex state, similarly to the experiment, \figref{fig3_Num}(d).

\paragraph{Weakly-nonlinear analysis.}
We examined the linear stability of Eq.~(\ref{eq:orig2}) around $\vb{v}=0$, yielding
$ \partial_t \omega = a \omega - (1+ \nabla^2)^2 \omega$.
Its solution is of the form $\omega=\sum_{-\infty}^\infty \exp(\lambda_n t) \omega_n$,
\begin{equation}
    \omega_n = \left(G_{n+} J_n (k_{n+} r) + G_{n-} J_n (k_{n-} r) \right)\exp (i n \theta),
    \label{sol1}
\end{equation}
where $\lambda_n$ are the growth rates of the corresponding azimuthal modes, $G_{n\pm} $ are constants, $J_n$ are the Bessel functions,
$k_{n\pm}=\sqrt{1\pm \sqrt{a-\lambda_n}}$. Applying the boundary conditions to $\omega_n$ and solving the characteristic equation (see Supplementary Note 4), one finds the growth rates $\lambda_n $ vs radius $R$. The results are shown in Fig.~\ref{fig4_Thm}(a).
For small enough $R$, all $\lambda_n$ are negative, so that no vortex is excited. For $R\gtrapprox 4.2$, $\lambda_0$ becomes positive, corresponding to the onset of the steady-state vortical motion observed in computational modeling and experiment, see \figref{S-fig:reff}(b) for quantitative agreement between the numerical and analytical solutions.
Then, with the gradual increase in $R$, higher rotational modes become unstable.
We find that vortex reversal occurs when the first two modes with $n=0, \pm1 $ are unstable, and the mode with $n=\pm 2$ is still stable but close to the threshold, compare Figs.~\ref{fig4_Thm}(a) and (e).

The radial vorticity and velocity profiles predicted by the linear analysis, Eq.~(\ref{sol1}), are in excellent agreement with the numerical solutions of Eq.~(\ref{eq:orig2}), Figs.~\ref{fig4_Thm}(b,c). Furthermore, fitting the theoretical expression, Eq.~(\ref{sol1}), for $n=0$ to experimental vorticity and velocity profiles of a stable vortex provides an excellent approximation as well, see Fig.~\ref{fig2_ExpData}(a) and Supplementary Note 1.

Next, we approximate the solution to Eq.~(\ref{eq:orig2}) as a sum of the three lowest azimuthal modes with $n=0,\pm1,\pm2$,
\begin{equation}
    \omega = C(t) \omega_0 (r) + [A_1(t)e^{i \theta}\omega_1(r) + A_2(t)e^{2 i \theta}\omega_2(r) + \mbox{c.c.}]
    \label{omega}
\end{equation}
Here $\omega_0, \omega_1, \omega_2$ are the eigenfunctions obtained from linear stability analysis. For definiteness, the eigenmodes are normalized by their kinetic energy, see Supplementary Note 4, Eqs.~(S35),(S36). $C(t), A_1(t), A_2(t)$ are slowly-varying amplitudes that are derived from the corresponding orthogonality leading to a set of normal form equations (\ref{C})-(\ref{A2}), see Methods.

Equations (\ref{C})-(\ref{A2}) faithfully reproduce the numerical results from Eq.~(\ref{eq:orig2}) without further approximation, see Fig.~\ref{fig4_Thm}. Specifically, for small radii, Eqs.~(\ref{C})-(\ref{A2}) reproduce a stable vortex solution as shown in Fig.~\ref{fig3_Num}(a). Then, with the increase in $R$, the bifurcation to a limit cycle is faithfully captured. Furthermore, even the details of the time dependence of each azimuthal mode closely agree with those of the numerical solutions of Eq.~(\ref{eq:orig2}), see Figs.~\ref{fig4_Thm}(d),~\ref{S-fig:reff}(c,d).
In the reversing vortex state, displayed in Fig.~\ref{fig4_Thm}(e), all three amplitudes $C, A_1, A_2$ are non-zero, see Supplementary Movie 14. With the further increase in $R$, a transition from a reversing state to a pulsating four-vortex solution occurs, see Fig.~\ref{fig4_Thm}(f) and Supplementary Movie 15. Here, the zero mode, $n=0$, is suppressed, and the first and second modes $A_1, A_2$ pulsate in anti-phase.
As shown in \figref{fig1_ExpSetup}(g), this anti-phase relation was indeed observed experimentally, further demonstrating the quantitative agreements among the experimental, numerical, and analytical results.
The normal form analysis indicates that the transition to vortex reversals and other time-dependent states is a result of resonant nonlinear interaction among the three lowest azimuthal modes.
This behavior only exists for sufficiently large values of the nonlinear advection term $\lambda \vb{v} \cdot \nabla \omega$ in Eq.~(\ref{eq:orig2}), which controls the resonant three-mode interaction. No limit cycles were found for $\lambda \lessapprox 3.75$.

\paragraph{Validating equations of motion.}
The use of the TTSHE was validated through regression analysis of our experimental data; see Supplementary Note 2 for details.
Similar approaches were used in Refs.~\cite{brunton2016discovering,supekar2023learning}.
In addition to the TTSHE, we tested another model for bacterial turbulence, the Nikolaevskiy equation, which includes $\nabla^6 \vb{v}$ term but no cubic nonlinearity $|\vb{v}|^2 \vb{v}$ nor linear term $\vb{v}$~\cite{beresnev1993model,tribelsky1996new,slomka2017spontaneous}.
The TTSHE outperformed the Nikolaevskiy equation in terms of the residuals, justifying our numerical and theoretical approaches, see Figs.~\ref{S-figS4_ParamEstNumerics},~\ref{S-figS5_ParamEstExp},~\ref{S-figS6_Residuals} and Tables.~\ref{S-table:TTSHparametersNumerics},~\ref{S-table:TTSHparametersExperiment},~\ref{S-table:Nikola}.
The regression for the two-vortex reversing state shown in \figref{fig1_ExpSetup}(f) yields $\lambda_\mathrm{dim} = 1.69 \pm 0.38$ for the dimensional TTSHE, proving the presence of the advection term with $\lambda > 1$, larger than $\lambda = 1$ for the Naiver-Stokes equation. Transforming the TTSHE into the form of Eq.~(\ref{eq:orig2}) with characteristic values in the unconstrained bacterial turbulence (velocity $V\approx$50~$\mathrm{\mu m/s}$, length scale $L\approx$40~$\mathrm{\mu m}$, and time scale $T\approx$0.5~s) yields $\lambda_\mathrm{nondim} = \frac{VT}{L} \lambda_\mathrm{dim} \approx 4.2$ in the dimensionless TTSHE\@. It is consistent with our theoretical prediction of $\lambda \gtrapprox 3.75$ for the onset of oscillations.

\subsection{Concluding remarks}
We observed a generic route to active turbulence in confined suspensions of swimming bacteria: a single steady vortex gives way to a reversing vortex pair, four pulsating vortices, and then to a well-developed spatiotemporal chaos. The fact that the entire bifurcation sequence is reproduced by a generic phenomenological model for active turbulence reveals the universal fundamental mechanism governing the transition: resonant interaction of the three lowest azimuthal modes associated with cylindrical confinement. Furthermore, the onset of the periodic reversal relies on the finite value of the Navier-Stokes-like advection term in the phenomenological model of active turbulence~\cite{wensink2012mesoscale,dunkel2013minimal,reinken2020organizing,shiratani2023route}.
The regression of experimental data also reliably corroborates the presence of the advection term with its coefficient $\lambda>1$ in the effective equation. These findings suggest that the observed transitions should also occur in a broad class of active self-propelled systems under confinement.
This robust mechanism is presumably responsible for the onset of reversing edge currents numerically observed in Ref.~\cite{matsukiyo2024oscillating} and is not sensitive to the details of boundary conditions or geometry~\cite{shiratani2023route}. Furthermore, the observed traditions occur in a Newtonian fluid environment. Viscoelasticity or anisotropy may only affect the details of the transition~\cite{liu2021viscoelastic,liao2023viscoelasticity,zhou2014living}. This generic mechanism is based on the three-mode resonant interaction and should be relevant for the variety of biological and synthetic active systems, e.g., Janus colloids~\cite{nishiguchi2015mesoscopic,iwasawa2021algebraic,nishiguchi2023deciphering}.

Another intriguing aspect is the effect of chirality. Since bacteria are chiral objects due to counter-rotation of the body and helical flagella~\cite{diluzio2005escherichia,marcos2012bacterial,jing2020chirality,grober2023unconventional}, there could be an asymmetry between CW/CCW rotating vortices~\cite{beppu2021edge}. In this work, a sustained effort was undertaken to make the upper/bottom surfaces of the wells as identical as possible to suppress the asymmetry. While a minor chiral shift does not affect the transition sequence, it could introduce slightly different thresholds for the
onsets of vortex oscillations of opposite chirality.

The current experiment is unavoidably susceptible to strong fluctuations discarded in the theoretical description. For example, the number of bacteria within a single microscopic well is about $\sim10^4$ bacteria/well. The dynamics of such a small bacterial population is intrinsically stochastic. Therefore, understanding how noise affects the nature of transitions and exploring ways to tame and control the fluctuating active dynamics would be of interest to future studies.

Finally, the controls and rectification of vortices in confined active matter open up new possibilities for engineering active matter.
For instance, weak coupling between neighboring wells may realize a ``bacterial lattice clock'', in which reversing vortex pairs synchronize and exhibit higher regularity and persistence. The reversing or pulsating vortices may be useful for mixing at low Reynolds numbers.
Taming the fluctuations in active systems based on the fundamental instability uncovered in this work provides novel design principles for functioning active devices, such as biosensors or microrobotic swarms for targeted drug delivery, precision surgery,
or detoxification~\cite{li2017micro,li20213d}.

\subsection{Methods}

\paragraph{Experimental details.} Bacteria \textit{Bacillus subtilis} (strain: 1085) were grown in Terrific Broth (T9179, Sigma-Aldrich) growth medium until optical density (OD) achieved $\mathrm{OD_{600\;nm}}\approx$1. After concentrating the suspension 180-fold, it was sandwiched between two thin polydimethylsiloxane (PDMS) membranes to facilitate sufficient oxygen supply for sustaining high bacterial motility. The bottom PDMS membrane was patterned with 30-$\mathrm{\mu m}$-deep multiple microscopic wells with the radius ranging from 44 to 51 $\mathrm{\mu m}$ with 0.5 $\mathrm{\mu m}$ increments [Figs.~\ref{fig1_ExpSetup}(a,b,d),~\ref{S-figS1_WholeFiledOfView},~\ref{S-figS2_DesignedDetectedRadius}]. To overcome systematic errors arising from different preparations of bacterial cultures and slight density variations caused during the confinement process, we simultaneously observed $\sim 400$ wells (19 radii, $\sim20$ wells for each radius) in a single field of view by using an inverted microscope equipped with a large-sensor sCMOS camera (Kinetix, Teledyne Photometrics, 3200$\times$3200 pixels) and a 10$\times$ objective lens, realizing the 2.1 mm $\times$ 2.1 mm field of view (\supfigref{S-figS1_WholeFiledOfView}). It allowed resolving the bacterial dynamics very close to the transition point.
We captured the movies at 50 fps for 150 seconds and analyzed the bacterial velocity fields $\vb{v}(\vb{r}, t)=(v_x, v_y)$ using the particle image velocimetry (PIV), see Supplementary Note 1 for the detailed protocols.

\paragraph{Computational details.}
Equation~\ref{eq:orig2} was solved by the pseudospectral method in a two-dimensional periodic $40.96 \times 40.96$ domain, discretized as the $8192 \times 8192$ square lattice.
Spatial derivatives were handled by the fast Fourier transform, see Supplementary Note 3.
Time update was performed in the Fourier space, with the time step $\Delta t = 0.01$.
To accelerate simulations, we performed the whole computation on GPUs (NVIDIA RTX A6000 or A100). 

The damping wall
implemented in Eq.~(\ref{eq:orig2}) with the kernel
$K (\vb{r})$
permits some leakage outside of the well radius $R$. We calibrated $R$ to account for the leakage
and defined the effective radius $R_{\mathrm{eff}}$ where the velocity and vorticity vanish.
$R_{\mathrm{eff}}$ is calculated as the root of $\int v_{\theta}\qty(r,\theta) \dd{\theta}$ (the zeroth azimuthal mode), Supplementary \figref{S-fig:reff}. We obtained $R_{\mathrm{eff}} -R \approx 0.5$.
The zeroth mode amplitude $C$ in~\eqref{omega} provides a convenient measure of the spin (\eqref{eq:spin}) up to a certain prefactor.
For the details of the numerical mode decomposition and related quantities, see Supplementary Note 4.

\paragraph{Normal form equations.} Substituting Eq.~(\ref{omega}) into Eq.~(\ref{eq:orig2}), and implementing the orthogonality conditions, we obtain the set of equations for amplitudes $C, A_1, A_2$
\begin{eqnarray}
    \partial_t C&=& \lambda_0 C - c_1 C^3 -c_2 C |A_1|^2 -c_3 C |A_2|^2
    \nonumber \\ &- & 2 c_4 \text{Re}A_2 A_1^{2*} \label{C} \\
    \partial_t A_1 & = & \lambda_1 A_1 - b_1 A_1 | A_1|^2 - b_2 A_1 C^2 - b_3 A_1 | A_2 |^2 \nonumber \\
    &-&b_4 C A_2 A_1^* +\delta_1 A_1 C +\gamma_1 A_2 A_1^* \label{A1} \\
    \partial_t A_2 & = & \lambda_2 A_2 - a_1 A_2 | A_2|^2 - a_2 A_2 C^2 - a_3 A_2 | A_1 |^2\nonumber \\
    &-&a_4 C A_1^2 + \delta_2 A_2 C +\gamma_2 A_1^2 \label{A2}
\end{eqnarray}
$\lambda_{0,1,2}$ are the linear growth rates; other coefficients are integrals over the nonlinearities, Supplementary Note 4.

\begin{acknowledgments}
    \paragraph{Acknowledgments.}
    This work is supported in part by JSPS KAKENHI Grant Numbers
    JP19H05800, 
    JP20K14426, 
    JP23K25838, 
    JP24K00593, 
    JP24KJ0890, 
    JST PRESTO Grant Number JPMJPR21O8, Japan, and
    the JSPS Core-to-Core Program ``Advanced core-to-core network for the physics of self-organizing active matter (JPJSCCA20230002)''. The research of I.S.A. was supported by the NSF award PHY-2140010.
    I.S.A. also acknowledges the visiting professorship for the Global Science Graduate Course -B, at The University of Tokyo, where main part of this work was done.
    S.S. acknowledges support from FoPM, WINGS Program, The University of Tokyo.
    The computation in this paper was done partly by the nekoya/ai cluster at the Institute for Physics of Intelligence, The University of Tokyo, and the facilities of the Supercomputer Center, the Institute for Solid State Physics, The University of Tokyo.

    \paragraph{Competing interests.}
    The authors declare no competing interests.

    \paragraph{Data availability.}
    All the experimental and numerical data, as well as the relevant codes and scripts, are available upon request.

    \paragraph{Author contributions.}
    D.N. conceived and initiated the project, performed the experiments, and analyzed the experimental data.
    S.S. performed the numerical simulations and analyzed the numerical data.
    I.S.A. developed the analytical theory. All authors discussed and interpreted the results, and wrote the manuscript.

\end{acknowledgments}

\bibliography{ref}

\end{document}


\title{Supplementary Material for\\ ``Vortex reversal is a precursor of confined bacterial turbulence''}

\author{Daiki Nishiguchi}
\email{nishiguchi@noneq.phys.s.u-tokyo.ac.jp}
\affiliation{Department of Physics,\! The University of Tokyo,\! 7--3--1 Hongo,\! Bunkyo-ku,\! Tokyo 113--0033,\! Japan}%

\author{Sora Shiratani}
\affiliation{Department of Physics,\! The University of Tokyo,\! 7--3--1 Hongo,\! Bunkyo-ku,\! Tokyo 113--0033,\! Japan}%

\author{Kazumasa A. Takeuchi}
\affiliation{Department of Physics,\! The University of Tokyo,\! 7--3--1 Hongo,\! Bunkyo-ku,\! Tokyo 113--0033,\! Japan}%
\affiliation{Institute for Physics of Intelligence, The University of Tokyo,\! 7--3--1 Hongo,\! Bunkyo-ku,\! Tokyo 113--0033,\! Japan}%

\author{Igor S. Aranson}
\affiliation{Departments of Biomedical Engineering, Chemistry, and Mathematics, The Pennsylvania State University, University Park, Pennsylvania 16802, USA}%
\affiliation{Department of Physics,\! The University of Tokyo,\! 7--3--1 Hongo,\! Bunkyo-ku,\! Tokyo 113--0033,\! Japan}%

\date{\today}

\maketitle

\section{Supplementary Note 1: Experimental details}
\subsection{Bacterial culture}
Bacteria \textit{Bacillus subtilis} (strain 1085) were first grown on an LB (Luria-Bertani) agar plate (BD Difco LB Broth Miller + 1.5 w\% agar) at $37^\circ$C. The grown bacterial colony was transferred into 13 mL of Terrific Broth (T9179, Sigma-Aldrich) medium in a glass test tube, sealed with a cap, and cultured at $37^\circ$C at 200 rpm. When it reached $\mathrm{OD_{600\;nm}}\approx$0.8--1.0, the test tube was taken out from the incubator and placed at room temperature ($23^\circ$C) for 45 minutes so that the bacteria became adjusted to the temperature before the experiments. The bacterial suspension was centrifuged at 3000 rpm (1057 rcf) for 2 minutes and concentrated 180-fold by removing the supernatant. The concentrated bacterial suspension was confined into the PDMS device for observation.

\subsection{Microfabrication}
The fabrication of the microfluidic device was performed with the standard soft lithography. The photoresist SU-8 3050 was spread on a silicon wafer at the thickness of 30 $\mathrm{\mu m}$, and circular patterns with multiple diameters were drawn with a maskless aligner $\mathrm{\mu}$MLA (Heidelberg Instruments) and developed following the standard protocol. Then, dimethylpolysiloxane (PDMS, Sylgard 184, Dow) was poured onto the patterned silicon wafer to form the layer of the thickness $\sim 500$~$\mathrm{\mu m}$.
We also poured PDMS onto another non-patterned bare silicon wafer to make a $\sim 300$-$\mathrm{\mu m}$-thick flat PDMS membrane to be used as the top plate.
The cured PDMS membranes were peeled off from the silicon wafers. The plasma cleaning was applied to the membranes to make the surface hydrophilic so that the bacterial suspensions can smoothly enter the small circular wells.

We placed the concentrated bacterial suspension on the patterned PDMS membrane and then sandwiched the suspension by placing the non-patterned PDMS membrane from the top. The use of the thin PDMS membranes for both top and bottom substrates realized sufficient oxygen supply to the dense bacterial suspension, enabling longer observation of highly active bacterial dynamics and the evaluation of vortex stability.

\begin{figure*}[t]
    \includegraphics[width=\hsize,clip]{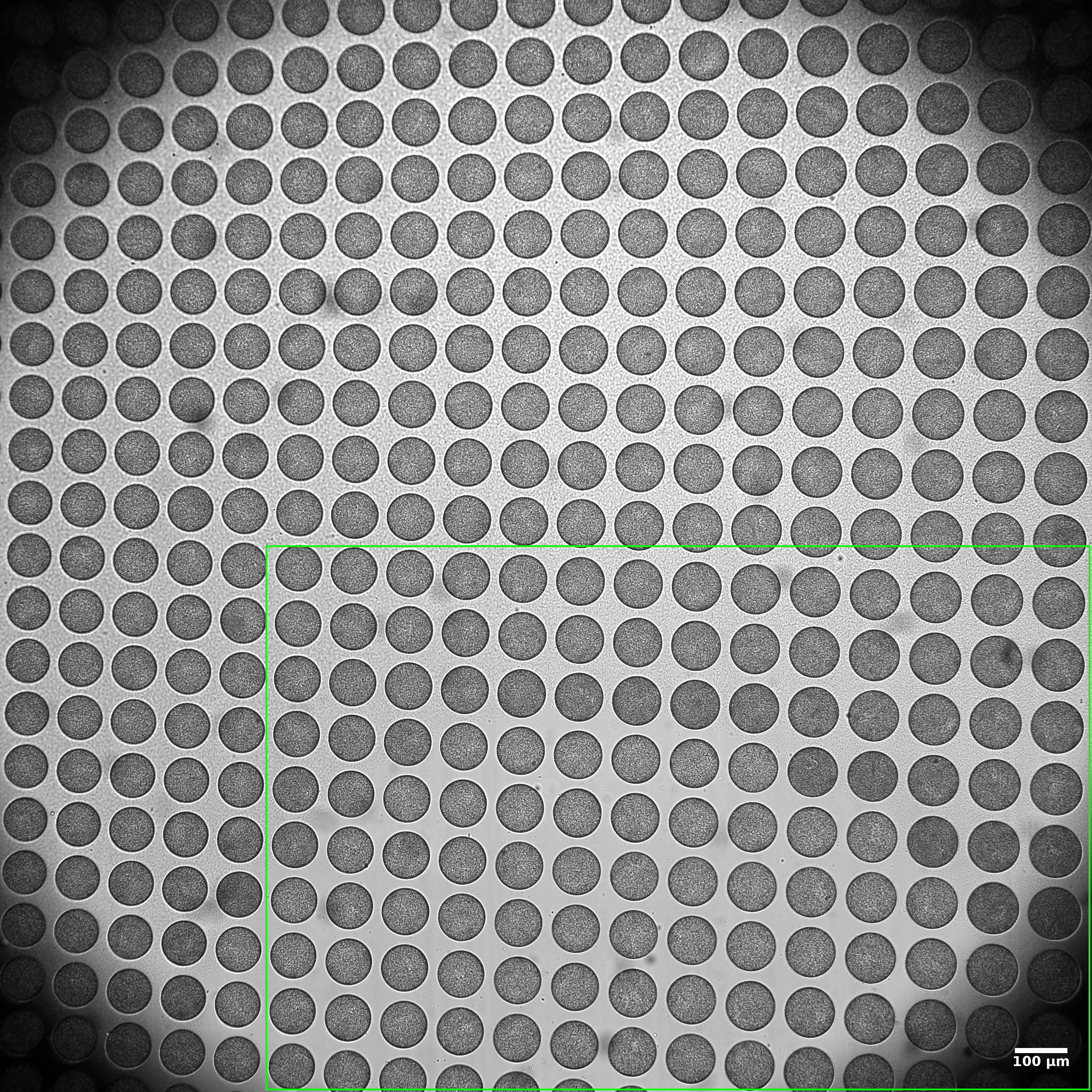}
    \centering
    \caption{%
        Snapshot of the whole field of view taken by the camera. The wells aligned in each column have the same designed radii and their radii are increasing from left to right. The wells whose centers were located within the green rectangular region were used for analysis. This rectangular region is displayed in Fig.1(c) in the main text.
    }\label{figS1_WholeFiledOfView}
\end{figure*}

\subsection{Microscopy}
The bright field images of the bacterial dynamics were captured with an inverted microscope (IX83, Evident) equipped with 10x objective lens (UPLSXAPO, NA=0.4) and a large-field-of-view sCMOS camera (Kinetix, Teledyne Photometrics, 3200$\times$3200 pixels). With this setup, we captureed a field of view of about 2.1 mm $\times$ 2.1 mm with the spatial resolution of 0.65 $\mathrm{\mu m}$/pixel at 50 fps for 150 seconds, enabling the simultaneous observation of multiple wells with different radii with 0.5 $\mathrm{\mu m}$ increment (\figref{figS1_WholeFiledOfView}).
This allowed us to avoid conducting multiple experiments for different radii using different bacterial cultures, thereby preventing uncontrolled variability in the sample states.

In addition, simultaneous observation of $\sim20$ wells with the same designed radii $R_\mathrm{design}$ was crucial for our purpose. The bacterial density within each well cannot be uniform due to the confinement process, which can be visually confirmed by the brightness of the bright field images (\figref{figS1_WholeFiledOfView}). The slight variations of the density largely affect the dynamics of the bacterial collective motion very close to the transition point. This is the reason why, even among the wells with the same designed radii $R_\mathrm{design}$, we observed variations of the spin correlation times [\figref{M-fig2_ExpData}(c)] and the CW-CCW bias in the rotational directions (\supfigref{figS3_CWbias}).

\subsection{Image processing}
The obtained movie was processed by using the open-source MATLAB plugin PIVlab~\cite{pivlab} to obtain the velocity field $\vb{v}_\mathrm{raw}(x, y, t)$. The PIV interrogation box size was 16$\times$16 pixels (10.4~$\mathrm{\mu m}$ $\times$10.4~$\mathrm{\mu m}$, sufficiently smaller than the length scale of bacterial collective motion) with the step size of 8 pixels (5.2~$\mathrm{\mu m}$, 50\% overlap).
Since the use of the PDMS membranes for both substrates instead of glass substrates resulted in the deflection of the device and reduced optical resolutions, we applied a three-dimensional Gaussian filter, $G(x, y, t) = \frac{1}{(2\pi)^{3/2}{\sigma_x} {\sigma_y} {\sigma_t}} \exp\left[-\left(\frac{x^2}{2{\sigma_x}^2}+\frac{y^2}{2{\sigma_y}^2}+\frac{t^2}{2{\sigma_t}^2}\right)\right]$, to the $\vb{v}_\mathrm{raw}(x, y, t)$ to remove the noise. The standard deviations of the Gaussian were chosen as $\sigma_x = \sigma_y = 16$~pixels (2 PIV grids) and $\sigma_t=0.04$~s (2 frames), which gave the best results in our experimental conditions. All the data presented in the main text were based on the filtered velocity field, $\vb{v}(x, y, t) = \iiint \vb{v}_\mathrm{raw}(x', y', t') G(x-x', y-y', t-t') \dd{x'} \dd{y'} \dd{t'}$.

The positions and the radii of the wells were detected by the circular Hough transform using the MATLAB function \texttt{imfindcircles}. A background image was first obtained by averaging all the 7500 frames of the movie. We applied an adaptive thresholding (contrast limited adaptive histogram equalization, CLAHE) to the background image to cancel the inhomogeneity of the illumination to ensure the accurate detection of the wells. The circular Hough transform was then applied to the processed image to detect the positions and the radii of the wells.

\begin{figure}[t]
    \includegraphics[width=0.6\hsize,clip]{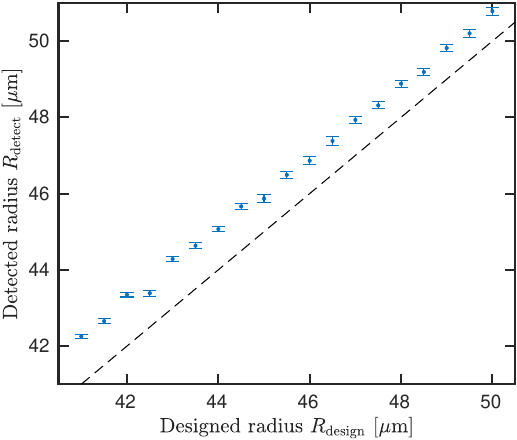}
    \centering
    \caption{%
        Comparison between the designed radii $R_\mathrm{design}$ and the detected radii $R_\mathrm{detect}$ of the wells. The wells in the field of view (\supfigref{figS1_WholeFiledOfView}) contained wells with $R_\mathrm{design}$ ranging from 41.0 to 50.0 $\mathrm{\mu m}$ with 0.5 $\mathrm{\mu m}$ increment. The detected radii $R_\mathrm{detect}$ are slightly larger than the designed radii $R_\mathrm{design}$, with the average difference of $0.98 \pm 0.05$~$\mathrm{\mu m}$ ($\pm$: standard errors).
        Blue circles are detection results. A black dashed line is the line $R_\mathrm{detect} = R_\mathrm{design}$ to guide the eye. Error bars: standard errors.
    }\label{figS2_DesignedDetectedRadius}
\end{figure}

\subsection{Analyzed region and bias in rotational directions}

Due to the confinement process, in some regions the bacterial suspensions were not completely confined within the wells and leaked out in between the top and bottom PDMS membranes, which we excluded from the analysis. Since the well patterns were fabricated only on the bottom PDMS membrane, the leakage breaks the up-down symmetry, resulting in weak bias in CW rotations. This is the reason for our choice of the green rectangular region used for the analysis (\figref{figS1_WholeFiledOfView}).

\begin{figure}[tb]
    \centering
    \includegraphics[width=0.6\hsize]{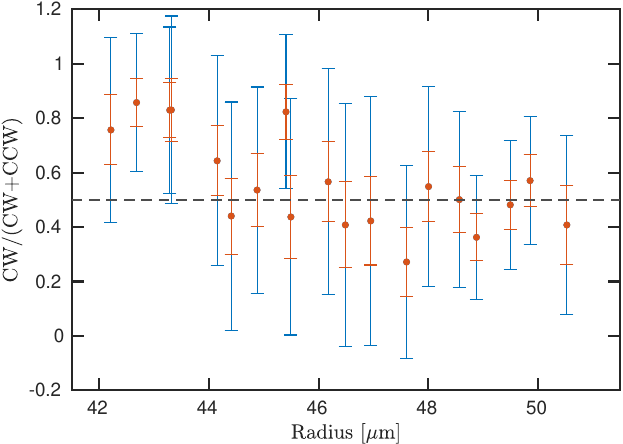}
    \caption{%
        The fraction of time that the spins rotated in the CW direction, $t_\mathrm{CW}/(t_\mathrm{CW}+t_\mathrm{CCW})$, as a function of the average detected radius $\langle R_\mathrm{detect} \rangle$ of the wells with the same designed radius $R_\mathrm{design}$. Error bars: standard deviations (blue) and standard errors (red). The wells with the radii $<44$~$\mathrm{\mu m}$ showed a significant bias in the CW direction, looking from above, due to the leakage of the bacterial suspension, which was excluded from the analysis.
    }\label{figS3_CWbias}
\end{figure}

The bias in CW rotation in the leaked regions looking from the above is exemplified in \figref{figS3_CWbias}, in which the wells with the smaller radii on the left side of the green rectangular region in \figref{figS1_WholeFiledOfView} are also plotted.
The bias in the rotational directions was quantified by calculating the fraction of time of CW rotation, $t_\mathrm{CW}/(t_\mathrm{CW}+t_\mathrm{CCW})$, where $t_\mathrm{CW}$ and $t_\mathrm{CCW}$ are the durations of CW and CCW rotations judged by the sign of the spin $S_i(t)$, respectively. We calculated this quantity for each well and then calculated the mean value of the wells with the same designed radius $R_\mathrm{design}$.
The absence of the CW-CCW bias was used as a criterion for determining the regions with good confinement to be used in the further analysis.

\subsection{Correlation time of the spins}
Correlation times of the spins were calculated to demonstrate the transition from the stabilized vortex state to the oscillatory state. For each well, we calculate the spin $S_i(t)$, where $i$ denotes the index of the well. Then, we calculated the correlation function,
\begin{align}
    C_i(\tau) = \frac{\langle S_i(t)S_i(t+\tau) \rangle_t}{\langle S_i(t)^2 \rangle_t},
\end{align}
where $\langle \;\rangle_t$ means temporal averaging. We defined the correlation time of the spin at which $C_i(\tau)$ becomes smaller than $1/e$ for the first time.
Within the experimental observation period of 150~s, some wells did not reverse the sign of $S_i(t)$ and exhibited quite stable behavior. Therefore, we could not define the correlation time for such wells and instead we plotted these wells at the vertical axis denoted as $>150$ in \figref{M-fig2_ExpData}(c). Since taking the average is not possible due to these data points with diverging correlation times, we calculated the moving median with the window width of 2~$\mathrm{\mu m}$. To highlight the discontinuity between the diverging data points at $>150$ and those with finite correlation times ($<150$), the red line connecting these two regions is plotted with a dashed line in Fig.2(c).

\subsection{Velocity \& vorticity profiles and fit to the analytical solution}
Quantitative comparisons between the experimental data and theoretical solutions were performed by examining the velocity and vorticity profiles for the single stabilized vortex, shown in \figref{M-fig2_ExpData}(a).
Due to the PIV interrogation box size 16$\times$16 pixels and the standard deviation of the Gaussian filter $\sigma_x = 16$~pixels, the obtained velocity
and vorticity fields can leak out from the radius $R$ detected by the image analysis by 32~pixels, corresponding to 20.8~$\mathrm{\mu m}$. This length scale is consistent with what is observed in \figref{M-fig2_ExpData}(a), where the experimental data extends beyond the radius $R$. These penetrating fields may justify to the numerical implementation of the damping terms for the velocity and vorticity fields.

The velocity and vorticity profiles in the single stabilized vortex, shown in~\figref{M-fig2_ExpData}(a), were fitted by the theoretical curves~\eqref{sol1} and~\eqref{vtheta_theory} for $n=0$.
The best fits yield $G_{0+} = 1.51$~$\mathrm{s}^{-1}$, $G_{0-} = 0.588$~$\mathrm{s}^{-1}$, $k_{0+} = 0.0914$~$\mathrm{\mu m}^{-1}$, $k_{0-} = 0.0453$~$\mathrm{\mu m}^{-1}$ for $\vb{v}$ and $G_{0+} = 1.31$~$\mathrm{s}^{-1}$, $G_{0-} = 0.633$~$\mathrm{s}^{-1}$, $k_{0+} = 0.0920$~$\mathrm{\mu m}^{-1}$, $k_{0-} = 0.0483$~$\mathrm{\mu m}^{-1}$ for $\omega$.

\section{Supplementary Note 2. Regression to the TTSHE}
\subsection{Least squares method}
We inferred the parameters in the TTSHE by regressing the experimental data to the TTSHE\@. Specifically, we applied the least squares method to the experimental data of the vorticity field $\omega(x, y, t)$ to find the parameters $A, C, \Gamma_0, \Gamma_2$ in the TTSHE\@. By assuming the vorticity equation for the TTSHE,
\begin{align}
    \lambda (\vb{v} \cdot \nabla) \omega + A \omega + C \nabla \times |\vb{v}|^2\vb{v} - \Gamma_0 \nabla^2 \omega + \Gamma_2 \nabla^4 \omega = - \partial_t \omega,
\end{align}
we constructed its matrix representation by using the experimental data. For a single regression, we used the data from time $t$ to $t+(N-1)\Delta t$ with $\Delta t$ being the time interval between the frames and $N$ being the number of frames used for the regression. The matrix representation is given by,
\begin{gather}
    \Phi
    P
    =
    \Psi,
\end{gather}
with,

\begin{gather}
    P = \begin{pmatrix}
        \lambda  \\
        A        \\
        C        \\
        \Gamma_0 \\
        \Gamma_2 \\
    \end{pmatrix},\;\;
    \Psi = \begin{pmatrix}
        -\partial_t \omega_t^{\vb{r}_1}                 \\
        -\partial_t \omega_t^{\vb{r}_2}                 \\
        \vdots                                          \\
        -\partial_t \omega_t^{\vb{r}_M}                 \\
        -\partial_t \omega_{t+\Delta t}^{\vb{r}_1}      \\
        -\partial_t \omega_{t+\Delta t}^{\vb{r}_2}      \\
        \vdots                                          \\
        -\partial_t \omega_{t+\Delta t}^{\vb{r}_M}      \\
        \vdots                                          \\
        -\partial_t \omega_{t+(N-1)\Delta t}^{\vb{r}_M} \\
    \end{pmatrix},
\end{gather}
\begin{align}
    \Phi = \begin{pmatrix}
               (\vb{v}_t^{\vb{r}_1} \cdot \nabla) \omega_t^{\vb{r}_1}                                 & \omega_t^{\vb{r}_1}                 & \nabla \times |\vb{v}_t^{\vb{r}_1}|^2 \vb{v}_t^{\vb{r}_1}                                 & -\nabla^2 \omega_t^{\vb{r}_1}                 & \nabla^4 \omega_t^{\vb{r}_1}                 \\
               (\vb{v}_t^{\vb{r}_2} \cdot \nabla) \omega_t^{\vb{r}_2}                                 & \omega_t^{\vb{r}_2}                 & \nabla \times |\vb{v}_t^{\vb{r}_2}|^2 \vb{v}_t^{\vb{r}_2}                                 & -\nabla^2 \omega_t^{\vb{r}_2}                 & \nabla^4 \omega_t^{\vb{r}_2}                 \\
               \vdots                                                                                 & \vdots                              & \vdots                                                                                    & \vdots                                        & \vdots                                       \\
               (\vb{v}_t^{\vb{r}_M} \cdot \nabla) \omega_t^{\vb{r}_M}                                 & \omega_t^{\vb{r}_M}                 & \nabla \times |\vb{v}_t^{\vb{r}_M}|^2 \vb{v}_t^{\vb{r}_M}                                 & -\nabla^2 \omega_t^{\vb{r}_M}                 & \nabla^4 \omega_t^{\vb{r}_M}                 \\
               (\vb{v}_{t+\Delta t}^{\vb{r}_1} \cdot \nabla) \omega_{t+\Delta t}^{\vb{r}_1}           & \omega_{t+\Delta t}^{\vb{r}_1}      & \nabla \times |\vb{v}_{t+\Delta t}^{\vb{r}_1}|^2 \vb{v}_{t+\Delta t}^{\vb{r}_1}           & -\nabla^2 \omega_{t+\Delta t}^{\vb{r}_1}      & \nabla^4 \omega_{t+\Delta t}^{\vb{r}_1}      \\
               (\vb{v}_{t+\Delta t}^{\vb{r}_2} \cdot \nabla) \omega_{t+\Delta t}^{\vb{r}_2}           & \omega_{t+\Delta t}^{\vb{r}_2}      & \nabla \times |\vb{v}_{t+\Delta t}^{\vb{r}_2}|^2 \vb{v}_{t+\Delta t}^{\vb{r}_2}           & -\nabla^2 \omega_{t+\Delta t}^{\vb{r}_2}      & \nabla^4 \omega_{t+\Delta t}^{\vb{r}_2}      \\
               \vdots                                                                                 & \vdots                              & \vdots                                                                                    & \vdots                                        & \vdots                                       \\
               (\vb{v}_{t+(N-1)\Delta t}^{\vb{r}_1} \cdot \nabla) \omega_{t+(N-1)\Delta t}^{\vb{r}_1} & \omega_{t+(N-1)\Delta t}^{\vb{r}_M} & \nabla \times |\vb{v}_{t+(N-1)\Delta t}^{\vb{r}_1}|^2 \vb{v}_{t+(N-1)\Delta t}^{\vb{r}_1} & -\nabla^2 \omega_{t+(N-1)\Delta t}^{\vb{r}_1} & \nabla^4 \omega_{t+(N-1)\Delta t}^{\vb{r}_1} \\
               \vdots                                                                                 & \vdots                              & \vdots                                                                                    & \vdots                                        & \vdots                                       \\
               (\vb{v}_{t+(N-1)\Delta t}^{\vb{r}_M} \cdot \nabla) \omega_{t+(N-1)\Delta t}^{\vb{r}_M} & \omega_{t+(N-1)\Delta t}^{\vb{r}_M} & \nabla \times |\vb{v}_{t+(N-1)\Delta t}^{\vb{r}_M}|^2 \vb{v}_{t+(N-1)\Delta t}^{\vb{r}_M} & -\nabla^2 \omega_{t+(N-1)\Delta t}^{\vb{r}_M} & \nabla^4 \omega_{t+(N-1)\Delta t}^{\vb{r}_M} \\
           \end{pmatrix},
\end{align}
where ${\vb{r}_1}$ to ${\vb{r}_M}$ denote the positions within the region of interest (ROI) used for the regression, and $\vb{v}_t^{\vb{r}_k} = \vb{v}({\vb{r}_k}, t)$ and $\omega_t^{\vb{r}_k} = \omega({\vb{r}_k}, t)$. Each column of $\Phi$ and $\Psi$ contains data from all the $M$ positions within the ROI and all the time points between $t$ and $t+(N-1)\Delta t$, totaling $MN$ elements.
This overdetermined system was solved by the least squares method using the pseudo-inverse matrix of $\Phi$ defined as,
\begin{align}
    \Phi^+ = (\Phi^T \Phi)^{-1} \Phi^T,
\end{align}
where $\Phi^T$ is the transpose of $\Phi$. Then, the optimal parameters $P_\mathrm{opt}$ were estimated as,
\begin{align}
    P_\mathrm{opt} = \Phi^+ \Psi.
\end{align}
Since the TTSHE is a model for the bulk flow, we excluded the region close to the confining boundary from the ROI so that the boundary does not affect the parameter estimations. The time window $N\Delta t$ was determined based on the spin correlation time. We repeated this regression at time $t$ and then shifted the time window to obtain the time series of estimated parameters. In this computation, we used central differences except for $\nabla^2\omega$ and $\nabla^4\omega$ that were calculated by using the fast Fourier transform.

\subsection{Verification with numerical data set}
To verify this method, we first applied this regression to the numerical data obtained from the TTSHE\@. We used the data for $R=5.30$, which shows the two oscillating vortices with the period of 117 in the numerical unit. A circular ROI centered at the well with half of the well's radius $R/2=2.65$ was chosen to diminish the boundary effects.
The time window was chosen to be $N\Delta t = 22$, which is the correlation time of this oscillatory state. To be consistent with experimental data, we decreased the spatial resolution of our high-resolution numerical data by 8 folds in both $x$ and $y$ directions, resulting in spatial resolution of $\Delta x = 0.32$, and $\Delta t$ was set to 1. These settings result in $M=217$ and $N=22$, and the duration of the numerical data was $500$. To estimate the standard errors of the average values, we used the effective number of independent samples as $N_\mathrm{sample} = 500/22 = 22.7$, which we calculated on the basis of the time window $N\Delta t = 22$~s and the duration of the data $500$.

\begin{figure}[tb]
    \centering
    \includegraphics[width=0.9\hsize]{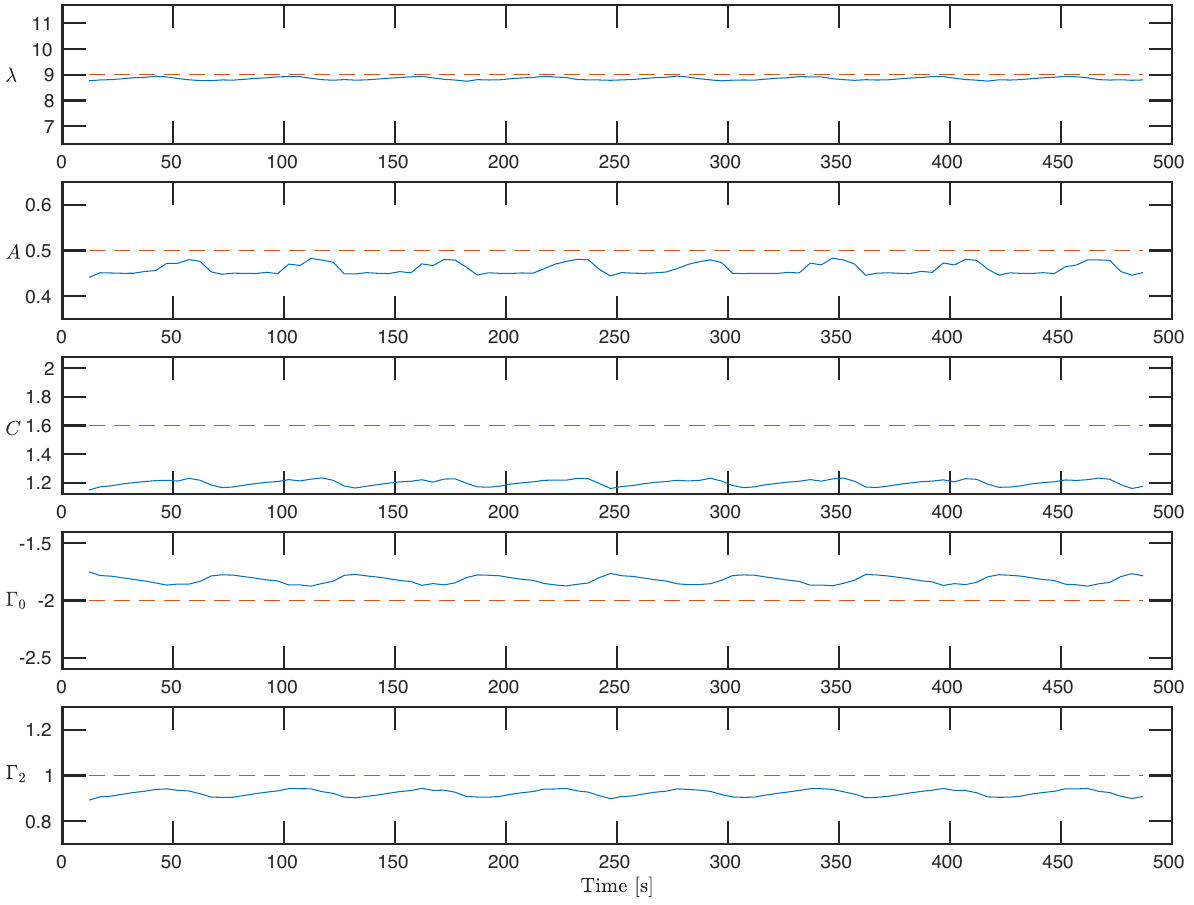}
    \caption{%
        Time series of the parameters in the TTSHE estimated from the regression of the numerical data. Blue curves are the results of the regression, and the red dashed lines are the ground truth values used for generating the numerical dataset. The vertical axes in all the plots ranges from -30\% to +30\% of the ground truth values.
    }\label{figS4_ParamEstNumerics}
\end{figure}

As shown in \supfigref{figS4_ParamEstNumerics} and in \tblref{table:TTSHparametersNumerics}, the time series of the estimated parameters fluctuate in time but their temporal averages gave reasonable values. Remarkably, the most accurate estimation was achieved for the most important parameter $\lambda$, which determines the strength of the nonlinear advection and the onset of the limit cycle. The error of the estimated $\lambda$ was as small as 2\%, while those of other parameters were 8\% to 25\%. Therefore, this estimation method can reliably be applied at least to estimating $\lambda$. The estimations of the other parameters have difficulties arising from their nonlinearities and higher-order derivatives.

\begin{table*}[tb]
    \caption{Estimated parameters of the TTSHE from the regression of the numerical data.}
    \begin{tabular}{c|c|c|c|c|c}
                           & $\lambda $ & $A$    & $C$    & $\Gamma_0$ & $\Gamma_2$ \\
        \hline
        Ground truth       & 9          & 0.5    & 1.6    & -2         & 1          \\Average & 8.844 & 0.4599 & 1.2033 & -1.8204 & 0.9231 \\
        Standard error     & 0.012      & 0.0026 & 0.0046 & 0.0073     & 0.0030     \\
        Standard deviation & 0.054      & 0.012  & 0.021  & 0.034      & 0.014

        \\
    \end{tabular}\label{table:TTSHparametersNumerics}
\end{table*}

\subsection{Application to experimental data}
Now we apply this regression to our experimental data.
We used the velocity and vorticity field data within the well with oscillatory behavior presented in \figref{M-fig1_ExpSetup}(e). We excluded the region within 32 pixels (20.8~$\mathrm{\mu m}$) from the boundary of the well, which composed the ROI for the regression.
The choice of 32 pixels was based on the sum of the PIV interrogation box size 16$\times$16 pixels and the standard deviation of the Gaussian filter $\sigma_x = 16$~pixels. Above this length scale, the estimated PIV velocity field is not affected by the boundary. With this choice, our ROI contained $M=107$ positions.

The time window $N\Delta t$ used for the regression was determined on the basis of the spin correlation time. Since the spin correlation time for this well was $14.2$~s, which corresponds to 710 frames, we chose $N=710$.

\begin{figure}[tb]
    \centering
    \includegraphics[width=0.6\hsize]{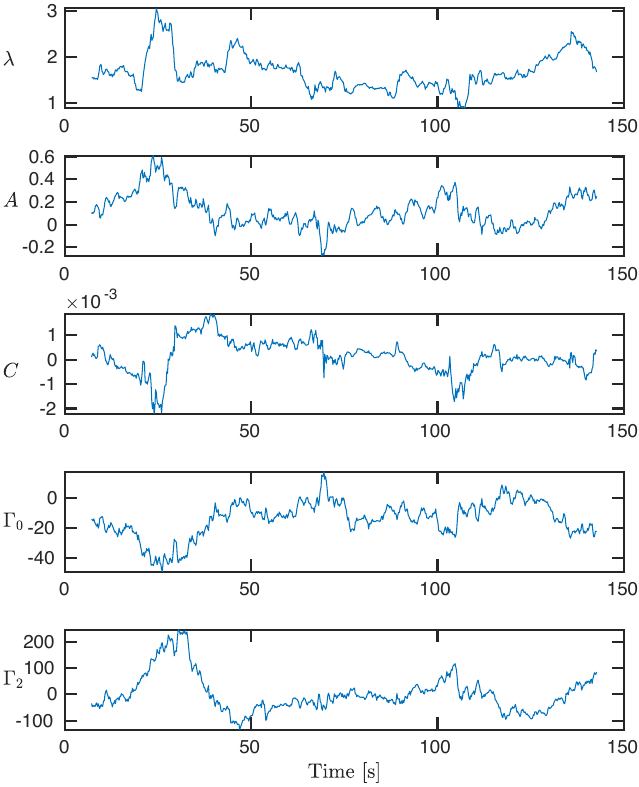}
    \caption{%
        Estimated parameters in the TTSHE from the regression of the experimental data as functions of time.
    }\label{figS5_ParamEstExp}
\end{figure}

We obtained the time series of estimated parameters shown in \supfigref{figS5_ParamEstExp}.
Based on the time window $N\Delta t = 14.2$~s and the duration of the movie 150~s, we estimate the effective number of independent samples as $N_\mathrm{sample} = 150/14.2 = 10.6$, which is then used for evaluating the standard errors of the estimated parameters. The average values and the standard errors of the estimated parameters are shown in \tblref{table:TTSHparametersExperiment}. As expected, the regression gave all the signs of the estimated parameters consistent with the numerical parameters in the main text.

Although the standard errors of the estimation were relatively large for the nonlinear term $C$ and the fourth-order derivative term $\Gamma_2$, the advection term $\lambda$ is rather reliably estimated, as expected from our control calculation with the numerical data set. The value $\lambda = 1.69 \pm 0.12$, which is larger than unity, is indeed a characteristic of the active turbulence model of pusher-type microswimmers such as \textit{Bacillus subtilis}. To compare with the nondimensional form of the TTSHE, we need to choose typical length $L$, time $T$ and velocity $V$. After nondimensionalization, all the coefficients scale as
$\lambda \rightarrow \frac{VT}{L} \lambda$,
$A \rightarrow T A$,
$C \rightarrow V^2 T C$,
$\Gamma_0 \rightarrow \frac{T}{L^2}\Gamma_0$, and
$\Gamma_2 \rightarrow \frac{T}{L^4}\Gamma_2$.
Therefore, if we choose a large value for $V$, e.g.\ based on an ideal unconstrained bacterial turbulence without confinement to the well, the value of $\lambda$ can be in the range where we found a limit cycle in our analytical calculation as we describe in the main text.

\begin{table*}[tb]
    \caption{Estimated parameters in the TTSHE from the regression of the experimental data. The brackets [ ] indicate the units of the parameters. The time-averaged values have signs consistent with our numerical parameters. Note that $\lambda$ is nondimensional.}
    \begin{tabular}{c|c|c|c|c|c}
                           & $\lambda $ & $A$ [s$^{-1}$] & $C$ [$\mathrm{\mu m^{-2}\cdot s}$] & $\Gamma_0$ [$\mathrm{\mu m^{2}\cdot s^{-1}}$] & $\Gamma_2$ [$\mathrm{\mu m^{4}\cdot s^{-1}}$] \\
        \hline
        Average            & 1.69           & 0.124          & 0.00011                            & -13.8                                         & 2.5                                           \\
        Standard error     & 0.12           & 0.047          & 0.00022                            & 3.9                                           & 23                                            \\
        Standard deviation & 0.38           & 0.14           & 0.00068                            & 11.9                                          & 71                                            \\
    \end{tabular}\label{table:TTSHparametersExperiment}
\end{table*}

\subsection{Regression to other equations}
The use of the TTSHE was justified on the basis of the regression to another model, Nikolaevskiy equation~\cite{beresnev1993model,slomka2017spontaneous}, also used as a model for the fluid motion of dense bacterial suspensions. Nikolaevskiy equation is, together with the incompressibility, given by,
\begin{gather}
    \nabla\cdot\vb{v}=0, \\
    \partial_t \vb{v} + (\vb{v}\cdot\nabla)\vb{v} = -\nabla p +
    \Gamma_0 \nabla^2 \vb{v} - \Gamma_2 \nabla^4 \vb{v} + \Gamma_4 \nabla^6 \vb{v}.
\end{gather}
We rewrote these equations into a vorticity equation by taking the rotation,
\begin{gather}
    \Gamma_0 \nabla^2 \omega - \Gamma_2 \nabla^4 \omega + \Gamma_4 \nabla^6 \omega =\partial_t \omega + (\vb{v}\cdot\nabla)\omega,
\end{gather}
and reconstructed a matrix representation for Nikolaevskiy equation for regression, in the same way as we did for the TTSHE\@. We obtained the optimal parameters as shown in \tblref{table:Nikola}, with physically reasonable signs, $\Gamma_0>0$, $\Gamma_2<0$, and $\Gamma_4>0$, for active turbulence with a characteristic length scale.

\begin{figure}[tb]
    \centering
    \includegraphics[width=0.9\hsize]{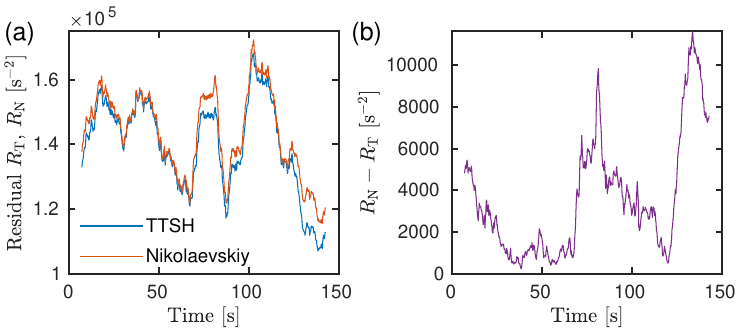}
    \caption{%
        Residuals of the regressions.
        (a) Time series of the residuals for TTSHE $R_\mathrm{T}$ (blue) and Nikolaevskiy equation $R_\mathrm{N}$ (red).
        (b) Time series of the difference of the residuals $R_\mathrm{N}-R_\mathrm{T}$.
    }\label{figS6_Residuals}
\end{figure}

We evaluated the performance of the two models by comparing the residuals of the regression, $R_\mathrm{T}$ for the TTSHE and $R_\mathrm{N}$
for Nikolaevskiy equation, respectively. Specifically, we calculated,
\begin{gather}
    R_\mathrm{T} = \| \Phi P_\mathrm{opt} - \Psi \|^2,
\end{gather}
where $\| \; \|$ denotes the norm of a vector. We compute these residuals for each time point $t$. The same procedure was applied also to Nikolaevskiy equation, and the time series of both models are plotted in \supfigref{figS6_Residuals}. As a result, $R_\mathrm{N}$ is always larger than $R_\mathrm{T}$, suggesting that the TTSHE outperforms Nikolaevskiy equation in terms of their ability to describe bacterial turbulence, at least, in our setup.

We note that it would be ideal to try modern machine learning techniques such as SINDy~\cite{brunton2016discovering} to infer the hydrodynamic description for bacterial turbulence by preparing the library of the possible terms.
However, this is beyond the scope of our current work.

\begin{table*}[tb]
    \caption{Estimated parameters in the Nikolaevskiy equation from the regression of the experimental data. The brackets [ ] indicate the units of the parameters.}
    \begin{tabular}{c|c|c|c}
                           & $\Gamma_0$ [$\mathrm{\mu m^{2}\cdot s^{-1}}$] & $\Gamma_2$ [$\mathrm{\mu m^{4}\cdot s^{-1}}$] & $\Gamma_4$ [$\mathrm{\mu m^{6}\cdot s^{-1}}$] \\
        \hline
        Average            & 11.5                                          & -803.7                                        & 2808.8                                        \\
        Standard error     & 2.6                                           & 135.0                                         & 662.5                                         \\
        Standard deviation & 8.0                                           & 417.4                                         & 2048.7

        \\
    \end{tabular}\label{table:Nikola}
\end{table*}

\section{Supplementary Note 3. Details of computational modeling}

\subsection{Model with Boundary}

To incorporate the effect of boundary, we use the following equation~\cite{wensink2012mesoscale,reinken2020organizing}:
\begin{equation}
    \begin{multlined}
        \pdv{\omega}{t} + \lambda \vb{v} \vdot \grad{\omega} = a \omega - b \curl{\qty[|\vb{v}|^2 \vb{v}]}
        - \qty(1 + \laplacian)^2 \omega - \gamma_{\vb{v}} \curl{\qty[K\qty(\vb{r})\vb{v}]} - \gamma_{\omega} K \qty(\vb{r}) \omega, \label{eq:ttsh_simple}
    \end{multlined}
\end{equation}
where $K\qty(\vb{r})$ is a positive-valued scalar field introduced to damp $\vb{v},\omega$ outside the well, and $\gamma_{\vb{v}},\gamma_{\omega} > 0$ are the associated damping strengths.
From its design, $K$ is defined to be $\simeq 0$ inside the well while $\simeq 1$ for the outside.
Similar to the model parameters, we also need to determine the values of $\gamma_{\vb{v}},\gamma_{\omega}$ based on experiments.
In this paper, we adopt the choice $\qty(\lambda,a,b,\gamma_{\vb{v}},\gamma_{\omega})=\qty(9,0.5,1.6,40,4)$ suggested in Ref.~\cite{reinken2020organizing}.

\subsection{Simulation Scheme}

To solve~\eqref{eq:ttsh_simple}, we use the pseudospectral method~\cite{reinken2020organizing}, which goes back and forth between the real and Fourier space to compute the spatial derivatives.
In this algorithm, the real space coordinates are defined on a periodic grid $\vb{r} \in \Delta x \cdot \mathbb{Z}_{\left[0,N-1\right]}^2$ where $N$ is the number of grid points per dimension and $\Delta x$ is the grid spacing.
For the Fourier space, wavenumber vector $\vb{k}$ resides in $\left( \pi / L \right) \cdot \qty[-N,N]^2$ where $L = N \Delta x$.
Each time we encounter spatial derivatives, we first Fourier transform quantities using the discrete Fourier transform (DFT), substitute $\grad$ with $i \vb{k} \cdot$, and then push the result back to the real space by the inverse DFT (iDFT).
The time evolution is done by the Euler method in the Fourier space since we can exactly handle the linear terms in a stable manner (operator splitting).
After updating $\omega$ (from time $t$ to $t + \Delta t$), we compute the associated incompressible velocity field $\vb{v}$ by the streamfunction method, which first solves the Poisson equation $\laplacian \psi = -\omega$ and then computes $\vu{v}_{x} = i \vb{k}_{y} \hat{\psi},\ \vu{v}_{y} = - i \vb{k}_x \hat{\psi}$ where $\hat{\ast}$ denotes the Fourier transform.
As this method does not fix the average velocity $\vu{v}\qty(\vb{0})$, we need to compute the dynamics of them separately.

In addition to the procedures described above, we need to take care of two simulation-specific issues.
First, we cannot compute the full dynamics for all the $\vb{k}$ because real-space multiplication doubles the range of $\vb{k}$.
To deal with that, we manually discard the high-$\vb{k}$ modes before and after real-space multiplications.
Similar treatment is also applied to $K$. Specifically, we use the following procedure to generate appropriate $K$ from the naive boolean mask $K_{\mathrm{naive}}\qty(\vb{r}) = 0\ \mathrm{if}\ \qty|\vb{r}| \leq R,\ 1 \ \mathrm{otherwise}$:
\begin{algorithm}[H]
    \begin{algorithmic}[1]
        \REQUIRE{$K_{\mathrm{naive}}$}
        \STATE{$\hat{K}_{\mathrm{naive}} = \dft{K_{\mathrm{naive}}}$}
        \FORALL{$\vb{k}$}
        \IF{$\vb{k}^2 > \qty(\frac{1}{4} \cdot \frac{\pi}{\Delta x})^2$}
        \STATE{$\hat{K}_{\mathrm{naive}} \qty(\vb{k}) = 0$}
        \ENDIF{}
        \ENDFOR{}
        \STATE{$K = \idft{\hat{K}_{\mathrm{naive}}}^2$}
        \ENSURE{processed $K(\vb{r})$}
    \end{algorithmic}
    \caption{smoothing $K$}\label{alg:mask}
\end{algorithm}
\noindent
Note that the positivity of $K$ is guaranteed by the square operation.
Second, numerical errors can lead to the violation of $\omega\qty(\vb{k}) = \omega\qty(-\vb{k})^\ast$, which is a necessary condition for the real-valued $\omega$.
To enforce this condition, we symmetrize $\omega$ each time after the Euler update.

Putting everything together, the simulation proceeds as follows:

\begin{algorithm}[H]
    \begin{algorithmic}[1]
        \REQUIRE{$\qty(\vb{v} \qty(t),\ \hat{\omega} \qty(t))$}
        \STATE{$\vb{rhs}_1 = -b \vb{v}^2 \vb{v} - \gamma_{\vb{v}} K \vb{v}$}
        \STATE{$\mathrm{rhs}_2 = -\lambda \vb{v} \vdot \idft{i \vb{k} \hat{\omega}} - \gamma_{\omega} K \omega$}
        \STATE{$\hat{\mathrm{rhs}} = i \vb{k} \times \vu{rhs}_1 + \hat{\mathrm{rhs}_2}$}
        \STATE{$\hat{\omega}_{\mathrm{naive}} = \exp\qty[\qty{a - \qty(1 - \vb{k}^2)^2}\Delta t]\qty(\hat{\omega} + \hat{\mathrm{rhs}} \Delta t)$}
        \FORALL{$\vb{k}$}
        \IF{$\vb{k} = \vb{0}$}
        \STATE{$\hat{\omega}_{\mathrm{new}} \qty(\vb{0}) = 0$}
        \STATE{$\vu{v}_{\mathrm{new}} \qty(\vb{0}) = \exp\qty[\qty(a - 1) \Delta t] \cdot \sum_{\vb{r}}\frac{1}{\sqrt{N^2}}(\vb{v} + \vb{rhs}_1 \Delta t)$}
        \ELSIF{$\vb{k}^2 > \qty(\frac{1}{2} \cdot \frac{\pi}{\Delta x})^2$}
        \STATE{$\hat{\omega}_{\mathrm{new}} \qty(\vb{k}) = 0$}
        \STATE{$\vu{v}_{\mathrm{new}} \qty(\vb{k}) = \vb{0}$}
        \ELSE{}
        \STATE{$\hat{\omega}_{\mathrm{new}}\qty(\vb{k})=\frac{1}{2}\qty{\hat{\omega}_{\mathrm{naive}}\qty(\vb{k})+\hat{\omega}_{\mathrm{naive}}\qty(\vb{-k})^*}$}
        \STATE{$\hat{\psi} \qty(\vb{k}) = \frac{\hat{\omega}_{\mathrm{new}} \qty(\vb{k})}{\vb{k}^2}$}
        \STATE{$\left. \vu{v}_{\mathrm{new}}\qty(\vb{k}) \right|_{x} = i \vb{k}_{y} \hat{\psi} \qty(\vb{k}),\ \left. \vu{v}_{\mathrm{new}} \qty(\vb{k}) \right|_{y} = - i \vb{k}_x \hat{\psi} \qty(\vb{k})$}
        \ENDIF{}
        \ENDFOR{}
        \ENSURE{$\qty(\vb{v}_{\mathrm{new}},\ \hat{\omega}_{\mathrm{new}}) = \qty(\vb{v} \qty(t + \Delta t),\ \hat{\omega} \qty(t + \Delta t))$}
    \end{algorithmic}
    \caption{simulation scheme}\label{alg:int}
\end{algorithm}
\noindent
In typical simulations, we set $N=8192$, $\Delta x=0.005$, and $\Delta t=0.01$.
All the computations were performed on GPUs (NVIDIA RTX A6000 or A100) in single precision, which achieves $\sim 60$ times speedup compared to CPU implementation with standard CPUs (e.g. Intel Xeon W-2295) \cite{shiratani2023route}.

\section{Supplementary Note 4. Analytical theory}
\subsection{Linear Theory}

We start with the TTSHE in the form
\begin{equation}
    \partial_t \omega + \lambda \vb{v}\cdot\nabla \omega = a \omega - (1+ \nabla^2)^2 \omega - b \nabla \times |\vb{v}|^2 \vb{v}.
    \label{she1}
\end{equation}
Consider a linearized Eq.~(\ref{she1}) in a disk domain of a radius $R$
\begin{equation}
    \partial_t \omega = a \omega - (1+ \nabla^2)^2 \omega.
    \label{she2}
\end{equation}
A generic solution to Eq.~(\ref{she2}) can be written in the form $\omega=\sum_{n=-\infty} ^\infty \omega_n$,
\begin{equation}
    \omega_n = \exp(\lambda_n t)\left(G_{n+} J_n (k_{n+} r) + G_{n-} J_n (k_{n-} r) \right)\exp (i n \theta),
    \label{sol1}
\end{equation}
where $\lambda_n$ is the growth rate, $G_{n\pm} $ arbitrary constants, $J_n$ are the Bessel functions, and $k_{n\pm}$ are given by the equation
\begin{equation}
    k_{n\pm }=\sqrt{1\pm \sqrt{a-\lambda_n}}.
    \label{k12}
\end{equation}

First, we calculate the stream function $\psi$ satisfying the condition
$\nabla^2 \psi = - \omega$. In polar coordinates, for each azimuthal mode $n$ we obtain
\begin{equation}
    \frac{1}{r} \partial_r (r \partial _r \psi_n)-\frac{n^2}{r^2} \psi _n = -\omega_n.
    \label{psi1}
\end{equation}
The solution is
\begin{equation}
    \psi _n = \left(\frac{G_{n+}}{k_{n+}^2}J_n(k_{n+} r)+ \frac{G_{n-}}{k_{n-}^2} J_n(k_{n-} r)+ G_{n0} r^n \right) \exp (i n \theta) + \mbox{c.c}.
    \label{psi2}
\end{equation}
Here, the term $G_{n0} r^n$ is a solution to the Laplace equation in polar coordinates $\frac{1}{r} \partial_r (r \partial _r \psi_n)-\frac{n^2}{r^2} \psi _n =0$.

Correspondingly, we obtain for the velocity components
\begin{align}
    v_\theta= - \partial _r \psi_n            & =- \left(\frac{G_{n+}}{2 k_{n+}}(J_{n-1}(k_{n+}k_ r)-J_{n+1}(k_{n+} r)) + \frac{G_{n-}}{2 k_{n-}}(J_{n-1}(k_{n-} r)-J_{n+1}(k_{n-} r)) + n G_{n0} r^{n-1} \right) \exp (i n \theta) + \mbox{c.c.}, \label{vtheta_theory} \\
    v_r = \frac{1}{r} \partial_\theta \psi _n & = \frac{i n}{r}\left(\frac{G_{n+}}{k_{n+}^2}J_n(k_{n+} r)+ \frac{G_{n-}}{k_{n-}^2} J_n(k_{n-} r)+ G_{n0} r^n \right) \exp (i n \theta) + \mbox{c.c.}
\end{align}
Now, to satisfy the b.c., we have the following conditions,
\begin{eqnarray}
    &&	G_{n+} J_n (k_{n+} R) + G_{n-} J_n (k_{n-} R) = 0, \\
    &&	\frac{G_{n+}}{k_{n+}^2}J_n(k_{n+} R)+ \frac{G_{n-}}{k_{n-}^2} J_n(k_{n-} R)+ G_{n0} R^n = 0, \\
    &&	\frac{G_{n+}}{2 k_{n+}}(J_{n-1}(k_{n+} R)-J_{n+1}(k_{n+} R)) + \frac{G_{n-}}{2 k_{n-}}(J_{n-1}(k_{n-} R)-J_{n+1}(k_{n-} R)) + n G_{n0} R^{n-1} =0.
\end{eqnarray}

We can set $G_{n+}=1$ due to system linearity. Then, the above 3 equations have 3 unknowns ($\lambda_n, G_{n-}, G_{n0}$). It provides the following characteristic equation to determine the growth rates $\lambda_n$ for arbitrary $n$:
\begin{equation}
    \frac{J_{n-1}(k_{n+} R)}{k_{n+} J_{n}(k_{n+} R)}-\frac{J_{n-1}(k_{n-} R)}{k_{n-} J_{n}(k_{n-} R)}- \frac{2 n}{R}(1/k_{n+}^2 -1/k_{n-}^2)=0
    \label{solv2}
\end{equation}

\subsection{Orthogonality condition}

Equation for the eigenmodes and eigenvalues
\begin{equation}
    \hat L \omega_n = (a - (1 +\nabla^2_n)^2)\omega_n = \lambda _n \omega _n
    \label{hatL}
\end{equation}
has rather subtle features (here $\nabla^2_n = r^{-1} \partial_r r \partial r - n^2/r^2 $ is the radial Laplacian). Namely, while the differential operator is symmetric, it is not formally self-adjoint since the orthogonality condition can not be applied to the vorticity modes $\omega_n $.
For the proper self-adjointness, the boundary conditions (b.c.) need to be explicit functions of the vorticity $\omega$ and its derivatives.
 Because part of the b.c. $\vb{v}=\vb{0}$ is expressed in terms of the velocity, the partial integration in the scaler product of the eigenfunctions leads to non-vanishing boundary terms.
By formulating the equations in terms of the stream function, the b.c can be reformulated in terms of the derivatives of the stream function, which leads to zero boundary terms in the scalar product of the eigenfunctions.
If we consider a scalar product,
\begin{equation}
    \langle \omega_n^k \omega_n^l \rangle \ne 0,
    \label{scalar}
\end{equation}the eigenmodes are not orthogonal.
Here $\omega_n^{l,k} $ are different modes corresponding to the same $n$. It can be seen from examining the product
$\langle \omega_n^k \hat L \omega_n^l \rangle $ that generates non-vanishing boundary terms after integration by parts. In the standard situation, the boundary terms vanish due to the boundary conditions on $\omega$. But we have instead boundary conditions for the stream function $\psi$.
It implies that the orthogonality should be implemented differently, for the stream function $\psi_n$ rather than the vorticity.

Therefore, we formulate the eigenvalue problem in terms of the stream function
\begin{equation}
    \hat L_1 \psi_n = (a - (1 +\nabla^2_n)^2) \nabla_n^2\psi_n = \lambda _n \nabla_n^2 \psi_n.
    \label{hatL1}
\end{equation}
The b.c.\ are $\psi_n =0, \partial_r \psi_n = 0 , \partial_r^2 \psi_n =0 $ for $r=R$.
With this b.c., the operator $\hat L_1$ is self-adjoint, and the boundary terms vanish identically. A little subtlety here is that it is a generalized eigenvalue problem because instead of $\lambda _n \psi_n$ we have $ \lambda _n \nabla_n^2 \psi_n $.

Now consider the orthogonality condition for each $\psi_n$ mode. The inhomogeneous equation is written as
\begin{equation}
    \hat L_1 w= (a - (1 +\nabla^2_n)^2) \nabla_n^2 w =-\partial_t \omega - \nabla \times \vb{B},
    \label{hatL2}
\end{equation}
where $\vb{B}= -b | \vb{v}| ^2\vb{v} - \lambda \vb{v} \cdot\nabla \vb{v}$. The solvability means that the r.h.s.\ is orthogonal to the zero eigenmodes $\psi_n$.
Applying the solvability condition, after partial integration we obtain
\begin{equation}
    \int_0^R d \theta r \psi^* \nabla \times \vb{B} dr = \int_0^R r \vb{v} ^*\vb{B} dr,
    \label{hatL2S}
\end{equation}
where $\vb{v} = (v_{r}, v_{\theta})$, where $v_r = \partial_\theta \psi /r , v_\theta =- \partial_r \psi $.
Correspondingly, for the normalization coefficients, we obtain, after partial integration,
\begin{equation}
    \int_0^R d \theta r \psi^* \omega dr = \int_0^R r |\vb{v}|^2 dr.
    \label{hat3}
\end{equation}

\subsection{Weakly-nonlinear analysis}
Here we consider a weakly-nonlinear solutions including only azimuthal modes with $n=0, \pm1, \pm 2 $,
\begin{equation}
    \psi = C(t) \psi_0 (r) + \left[A_1(t)\exp(i \theta)\psi_1 + A_2(t)\exp(2 i \theta)\psi_2+\mbox{c.c.}\right] + w.
    \label{nonl1}
\end{equation}
Here $\psi_0, \psi_1, \psi_2$ are the eigenfunctions obtained from linear stability analysis, $C(t), A
        _{1,2}(t)$ are slowly-varying functions, and $w$ is small
correction to the solution. Functions $C(t), A_1(t), A_2(t)$ are obtained from the corresponding orthogonality conditions guaranteeing that $w$ does not grow.

Substituting solution Eq.~(\ref{nonl1}) into Eq.~(\ref{she1}), and retaining terms for each azimuthal harmonics, we obtain after applying the orthogonality conditions,
\begin{eqnarray}
    \partial_t C&=& \lambda_0 C - c_1 C^3 -c_2 C |A_1|^2 -c_3 C |A_2|^2 - 2 c_4 \text{Re}A_2 A_1^{2*} ,\label{C} \\
    \partial_t A_1 & = & \lambda_1 A_1 - b_1 A_1 | A_1|^2 - b_2 A_1 C^2 - b_3 A_1 | A_2 |^2-b_4 C A_2 A_1^* + \delta_1 A_1 C +\gamma_1 A_2 A_1^* ,\label{A1} \\
    \partial_t A_2 & = & \lambda_2 A_2 - a_1 A_2 | A_2|^2 - a_2 A_2 C^2 - a_3 A_2 | A_1 |^2-a_4 C A_1^2 + \delta_2 A_2 C +\gamma_2. A_1^2 \label{A2}
\end{eqnarray}
Here $\lambda_{0,1,2} $ are the linear growth rates, and all other coefficients are integrals over the eigenfunctions.
In the following, we can consider $A_{1,2}$ complex.

\subsection{Calculations of the nonlinear coefficients}
We substitute the approximate solution Eq (\ref{nonl1}) into equation (\ref{she1}) and apply the solvability conditions.
The coefficients $m_{0,1,2}$ are the integrals of the eigenmode squares
\begin{equation}
    m_{0,1,2}= \int_0^R d r r |\vb{v} _{0,1,2}(r)|^2 ,\label{eq:m012}
\end{equation}
where the normalized radial eigenmodes are given by Eq.~(\ref{psi2})
\begin{equation}
    \psi _n = \frac{1}{\sqrt{m_n}}\left(\frac{G_{n+}}{k_{n+}^2}J_n(k_{n+} r)+ \frac{G_{n-}}{k_{n-}^2} J_n(k_{n-} r)+ G_{n0} r^n \right).
\end{equation}

The nonlinear terms in Eqs.~(\ref{C}), (\ref{A1}), (\ref{A2}) appear from the quadratic term $\vb{v} \cdot \nabla \omega$ and the cubic term $\nabla \times |\vb{v}|^2 \vb{v} $.

The coefficients for the quadratic terms $\delta_{1,2}, \gamma_{1,2}$ are given by the expressions (after partial integration)
\begin{eqnarray}
    \delta_1 A_1 C +\gamma_1 A_2 A_1^*	 & =& - \frac{\lambda}{2 \pi} \int d \theta d r r \vb{v} _1^* \exp(-i \theta) \vb{v} \cdot \nabla \vb{v}, \\
    \delta_2 A_2 C +\gamma_2 A_1^2	 & =& - \frac{\lambda}{2 \pi} \int d \theta d r r \vb{v} _2^* \exp(-2 i \theta) \vb{v} \cdot \nabla \vb{v}. \\
\end{eqnarray}
Here $\vb{v}_n = (v_{rn}, v_{\theta n})= (i n \psi_n/r, - \partial_r \psi_n)$. Note that in polar coordinates,
\begin{equation}
    \vb{v} \cdot \nabla \vb{v}= (v_r \partial_r v_r + v_\theta \partial_\theta v_r - v_\theta^2/r, v_r \partial_r v_\theta + v_\theta \partial_\theta v_\theta+ v_\theta v_r/r).
\end{equation}

The advection term does not generate any contribution to Eq.~(\ref{C}) due to the symmetry.

Correspondingly, for $a_{1,2,3,4}, b_{1,2,3,4}, c_{1,2,3,4}$ we obtain
\begin{eqnarray}
    c_1 C^3 +c_2 C |A_1|^2 +c_3 C |A_2|^2 +2c_4 \mathrm{Re}A_2 A_1^{2*} & =& \frac{b}{2 \pi} \int d \theta d r r \vb{v}_0^* \vb{v} |\vb{v}|^2, \\
    b_1 A_1 | A_1|^2 + b_2 A_1 C^2 + b_3 A_1 | A_2 |^2 + b_4 C A_2 A_1^* & =& \frac{b}{2 \pi} \int d \theta d r r \vb{v}_1^* \exp(-i \theta) \vb{v} |\vb{v}|^2, \\
    a_1 A_2 | A_2|^2 + a_2 A_2 C^2 + a_3 A_2 | A_1 |^2 +a_4 C A_1^2& =& \frac{b}{2 \pi} \int d \theta d r r \vb{v}_2^* \exp(-2 i \theta) \vb{v}
    |\vb{v}|^2.
\end{eqnarray}

All linear and nonlinear coefficients are calculated in Mathematica
and imported directly into the normal form equations (\ref{C}), (\ref{A1}), (\ref{A2}). It appears that $\gamma_1=\gamma_2$, and $c_4 =a_4 = b_4/2$.
The resulting equations generate a limit cycle without any further adjustments.

\subsection{Bridging theory and numerics}
\begin{figure}
    \centering
    \begin{minipage}[t]{0.45\hsize}
        \raggedright{{\large (a)}}
        \includegraphics[width=\hsize]{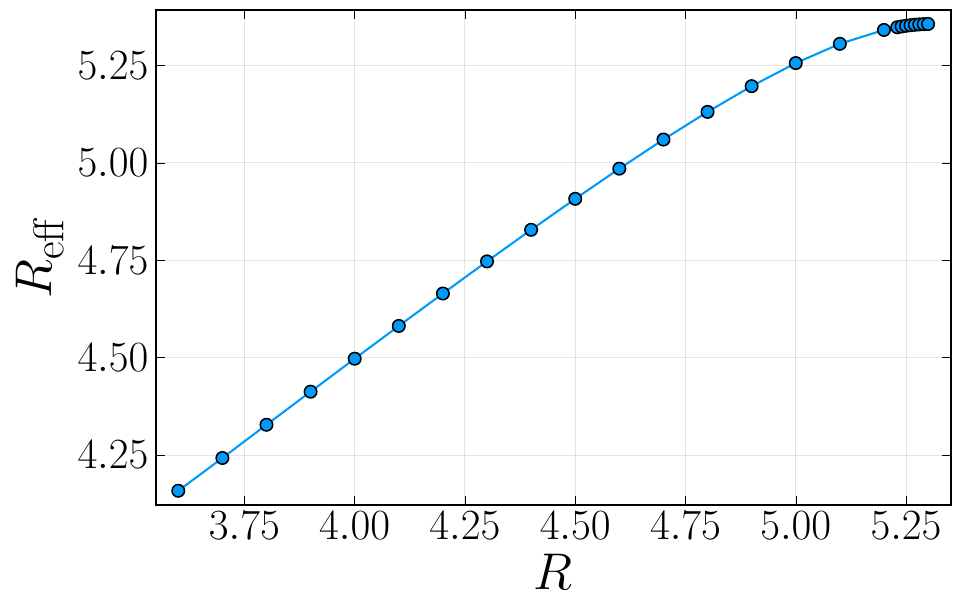}
    \end{minipage}
    \begin{minipage}[t]{0.45\hsize}
        \raggedright{{\large (b)}}
        \includegraphics[width=\hsize]{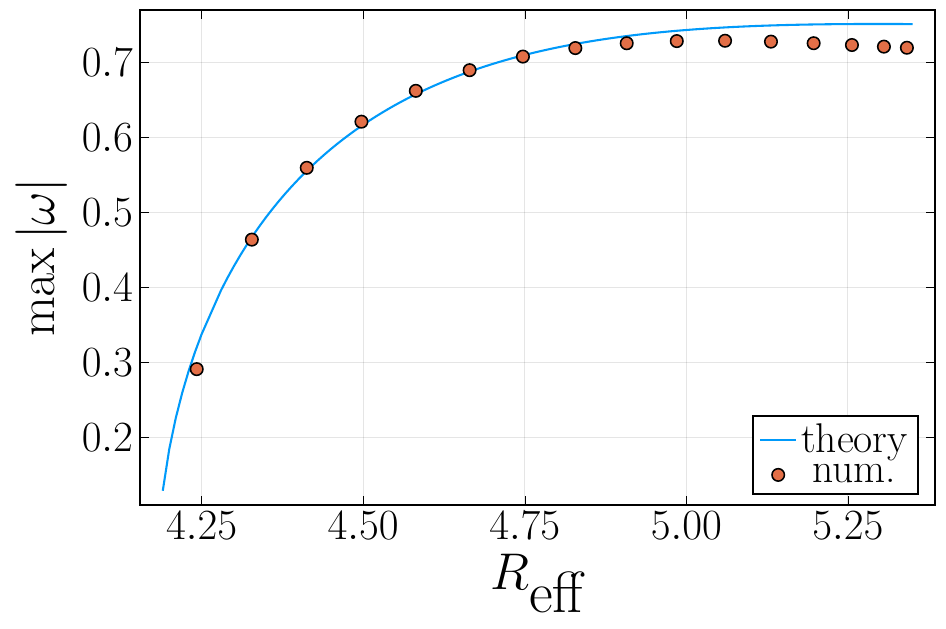}
    \end{minipage}
    \begin{minipage}[t]{0.45\hsize}
        \raggedright{{\large (c)}}
        \includegraphics[width=\hsize]{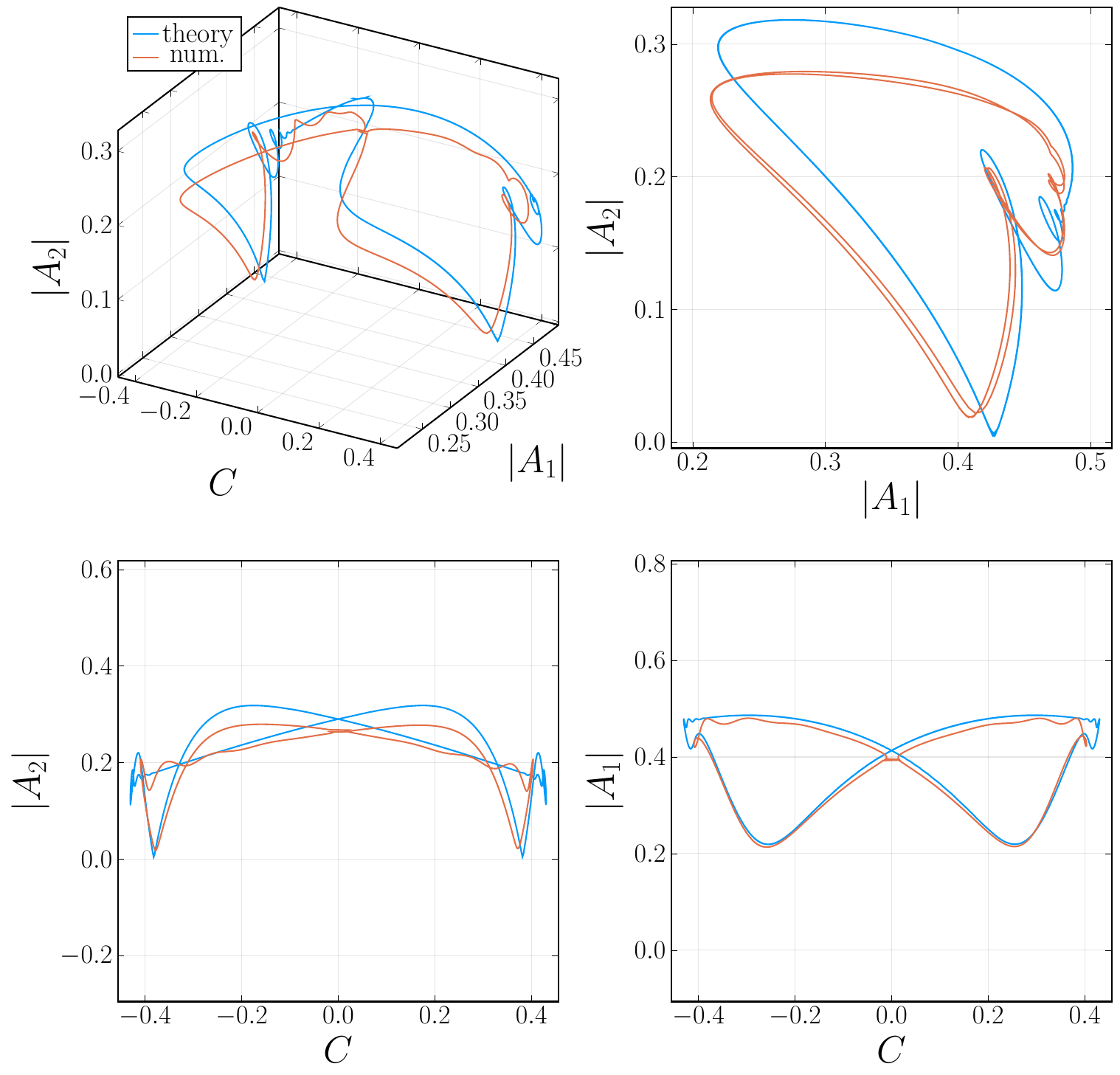}
    \end{minipage}
    \begin{minipage}[t]{0.45\hsize}
        \raggedright{{\large (d)}}
        \includegraphics[width=\hsize]{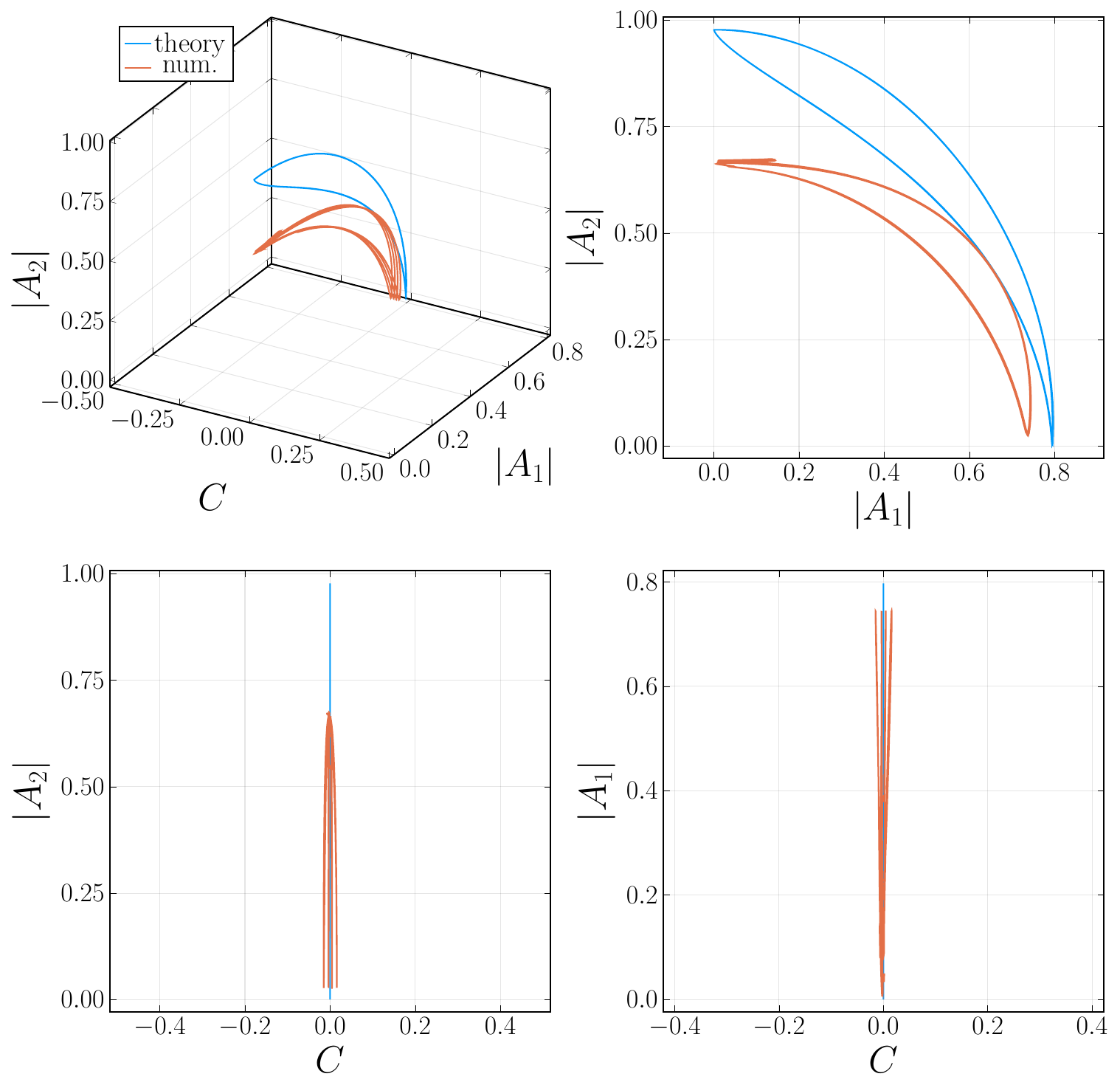}
    \end{minipage}

    \caption{%
        (a) Radius $R$ used for $K$ (damping mask) and effective radius $R_{\mathrm{eff}}$ estimated from the simulation.
        Generally we have $R_{\mathrm{eff}} > R$ due to leakage.
        (b) \(R\)-dependence of \(\max \left|\omega\right|\) near the lower bound radius having $\lambda_0 > 0$.
        (c,d) Three-dimensional phase space plot and projections, in the two-vortex state ((c), $R=5.35$ for simulation and $R=5.9$ for theory) and four-vortex state ((d), $R=6.4$ for simulation and $R=7.0$ for theory).
    }\label{fig:reff}
\end{figure}

To perform faithful comparison, mode decomposition of numerical data was done by spline interpolation on the radial grid followed by the trapezoidal quadrature based on the same formula as in the analytical theory.
The upper bound of $r$ integral was set larger than $R$ (typically $\simeq 9$) to contain the well plus the leakage.
The azimuthal modes (\figref{M-fig3_Num}(c), \figref{M-fig4_Thm}(c,d)) were based on the following formulae:
\begin{equation}
    v_{\ast n} = \frac{1}{2 \pi \sqrt{m_n}} \int \dd{\theta} e^{-i n \theta} v_{\ast} \qty(r,\theta),
\end{equation}
where \(\ast = r,\theta\).
Similarly the associated mode amplitudes (\figref{M-fig3_Num}(b), \figref{M-fig4_Thm}(b)) were calculated by the following:
\begin{gather}
    \qty|C| = \sqrt{\frac{\int \dd{r} r\qty|v_{\theta 0, \mathrm{num.}}|^2}{\int \dd{r} r\qty|v_{\theta 0, \mathrm{theory}}|^2}},\\
    \qty|A_1| = \sqrt{\frac{\int \dd{r} r\qty|v_{\theta 1, \mathrm{num.}}|^2}{\int \dd{r} r\qty|v_{\theta 1, \mathrm{theory}}|^2}},\\
    \qty|A_2| = \sqrt{\frac{\int \dd{r} r\qty|v_{\theta 2, \mathrm{num.}}|^2}{\int \dd{r} r\qty|v_{\theta 2, \mathrm{theory}}|^2}}.
\end{gather}
As for the sign of $C$, as $v_{\theta 0, \mathrm{theory}} > 0$, we can naturally identify it as the sign of $\int \dd{r} v_{\theta 0, \mathrm{num.}}$.
Moreover, beyond the sign consistency, the spin variable defined in \eqref{M-eq:spin} is proportional to $C$, as readily verified: 
$S_i \propto \iint r \dd{r} \dd{\theta} r v_\theta = \int \dd{r} r^2 \int \dd {\theta} \qty[C v_{\theta 0} + \sum_{n \geq 1} \qty(A_n v_{\theta n} e^{i n \theta} + \mathrm{c.c.})] = C \int \dd{r} r^2 v_{\theta 0}  \propto C$,
where $A_n$ are the mode amplitudes and the integrals run over the area of the $i$-th well. This proportionality was used in \figref{M-fig3_Num}(b).
With the help of mode decomposition and $R_{\mathrm{eff}}$ plotted in \figref{fig:reff}(a), we can almost directly compare the theory and simulation for some quantities, such as $\max \qty|\omega|$ shown in \figref{fig:reff}(b).

In the large-$R$ region, where we cannot compute reliable estimates of $R_{\textrm{eff}}$ due to oscillation, we still see good agreements between the theory and simulation as in \figref{fig:reff}(c,d).

\section{Supplementary Movie captions}

\begin{description}
    \item[Movie 1]
          Movie of the whole field of view (2.1~mm $\times$ 2.1~mm), corresponding to \figref{figS1_WholeFiledOfView}. The initial 20 seconds of the data are presented at twice the real speed ($2\times$).
          The spatial resolution is lowered to reduce the file size.
    \item[Movie 2]
          The vorticity and velocity fields of the single stabilized vortex ($R=44.6$~$\mathrm{\mu m}$). The color bar of the vorticity field is the same as in \figref{M-fig1_ExpSetup}. The movie is played at real-time speed. The scale bar represents 20 $\mathrm{\mu m}$.
    \item[Movie 3]
          The vorticity and velocity fields of the reversing vortices ($R=46.7$~$\mathrm{\mu m}$). The color bar of the vorticity field is the same as in \figref{M-fig1_ExpSetup}. The movie is played at real-time speed. The scale bar represents 20 $\mathrm{\mu m}$.
    \item[Movie 4]
          The vorticity and velocity fields of the four-vortex state ($R=48.8$~$\mathrm{\mu m}$). The color bar of the vorticity field is the same as in \figref{M-fig1_ExpSetup}. The movie is played at real-time speed. The scale bar represents 20 $\mathrm{\mu m}$.

    \item[Movie 5]
          The velocity field of the single stabilized vortex ($R=44.6$~$\mathrm{\mu m}$).
          The velocity vectors are colored based on the sign of the local vorticity: yellow for positive and green for negative. The movie is played at $4\times$ real-time speed. The scale bar represents 20 $\mathrm{\mu m}$.
    \item[Movie 6]
          The velocity field of the reversing vortices ($R=46.7$~$\mathrm{\mu m}$). The velocity vectors are colored based on the sign of the local vorticity: yellow for positive and green for negative. The movie is played at $4\times$ real-time speed. The scale bar represents 20 $\mathrm{\mu m}$.
    \item[Movie 7]
          The velocity field of the four-vortex state ($R=48.8$~$\mathrm{\mu m}$). The velocity vectors are colored based on the sign of the local vorticity: yellow for positive and green for negative. The movie is played at $4\times$ real-time speed. The scale bar represents 20 $\mathrm{\mu m}$.
    \item[Movie 8]
          Vorticity field of the single stabilized vortex (\figref{M-fig3_Num}(a), left), numerically computed at $R=5.2$.
    \item[Movie 9]
          Vorticity field of the reversing two-vortex state (\figref{M-fig3_Num}(a), middle), numerically computed at $R=5.6$.
    \item[Movie 10]
          Vorticity field of the pulsating four-vortex state (\figref{M-fig3_Num}(a), right), numerically computed at $R=6.4$.
    \item[Movie 11]
          Vorticity field of the reversing two-vortex state(\figref{M-fig3_Num}(b)), numerically computed at $R=5.35$.
    \item[Movie 12]
          Azimuthal Fourier decomposition of the vorticity field of the reversing two-vortex state at $R=5.6$ (\figref{M-fig3_Num}(c)). The original vorticity field and the modes $n=0, 1, 2$ are shown.
    \item[Movie 13]
          Vorticity field of the pulsating four-vortex state (\figref{M-fig3_Num}(d)), numerically computed at $R=6.2$.
    \item[Movie 14]
          Vorticity field and its azimuthal Fourier decomposition of the analytically calculated reversing two-vortex state at $R=5.9$ (\figref{M-fig4_Thm}(e)). The definitions of the modes are the same as in Supplementary Movie 12 and \figref{M-fig3_Num}(c).
    \item[Movie 15]
          Vorticity field and its azimuthal Fourier decomposition of the analytically calculated pulsating four-vortex state at $R=7.0$ (\figref{M-fig4_Thm}(f)). The definitions of the modes are the same as in Supplementary Movie 12 and \figref{M-fig3_Num}(c).
\end{description}

\bibliography{ref}